\documentclass[12pt]{iopart}
\usepackage{epsfig}
\usepackage{graphicx}

\newcommand{\be}{\begin{equation}}
\newcommand{\ee}{\end{equation}}
\newcommand{\bd}{\begin{displaymath}}
\newcommand{\ed}{\end{displaymath}}

\newcommand{\pprime}{{\prime\prime}}
\newcommand{\bra}{\langle}
\newcommand{\ket}{\rangle}
\newcommand{\order}{{\cal O}}
\newcommand{\smallo}{{o}}

\newcommand{\minus}{{\!-\!}}
\newcommand{\sgn}{{\rm sgn}}
\newcommand{\erf}{{\rm erf}}
\newcommand{\openone}{{\rm 1\!\!I}}

\newcommand{\bq}{\ensuremath{\mathbf{q}}}

\newcommand{\bw}{\ensuremath{\mathbf{w}}}
\newcommand{\bx}{\ensuremath{\mathbf{x}}}

\newcommand{\bz}{\ensuremath{\mathbf{z}}}

\newcommand{\bR}{\ensuremath{\mathbf{R}}}

\newcommand{\hq}{\hat{q}}
\newcommand{\hw}{\hat{w}}
\newcommand{\hx}{\hat{x}}

\newcommand{\hA}{\hat{A}}
\newcommand{\hC}{\hat{C}}

\newcommand{\hK}{\hat{K}}
\newcommand{\hL}{\hat{L}}
\newcommand{\D}{{\cal D}}

\newcommand{\hOmega}{\hat{\mbox{$\Omega$}}}
\newcommand{\bxi}{{\mbox{\boldmath $\xi$}}}
\newcommand{\bomega}{{\mbox{\boldmath $\omega$}}}
\newcommand{\bpsi}{{\mbox{\boldmath $\psi$}}}

\newcommand{\bOmega}{{\mbox{\boldmath $\Omega$}}}
\newcommand{\rate}{\tilde{\eta}}

\begin{document}

\title[Dynamical Solution of the On-Line Minority Game]{Dynamical Solution of the On-Line Minority Game}
\author{
A C C Coolen and J A F Heimel}
\address{
Department of Mathematics, King's College London\\ The Strand,
London WC2R 2LS, UK }

\begin{abstract}
We solve the dynamics of the on-line minority game, with general
types of decision noise, using
 generating functional techniques a la De Dominicis
and the temporal regularization  procedure of Bedeaux et al.
 The result is a macroscopic dynamical theory
  in the form of closed equations for correlation- and response
functions defined via an effective continuous-time single-trader
process, which are exact in both the ergodic and in the
non-ergodic regime of the minority game. Our solution also
explains why, although one cannot formally truncate the
Kramers-Moyal expansion of the process after the Fokker-Planck
term, upon doing so one still finds the correct solution, that the
previously proposed diffusion matrices for the Fokker-Planck term
are incomplete, and how previously proposed approximations of the
market volatility can be traced back to ergodicity assumptions.
\end{abstract}


\section{Introduction}

The Minority Game (MG) \cite{ChalZhan97} is an intriguing
variation on the so-called El-Farol bar problem \cite{Arth94}
which aims to capture and understand in the simplest possible way
the cooperative phenomena in markets of interacting traders. Its
equations describe the stochastic evolution in time of the
selection of individual `trading strategies' by a community of
traders which operate in a simple market. The rules of this market
are that each trader has to make a binary decision at every point
in time (e.g. whether to buy or sell), and that profit is made
only by those traders who find themselves in the minority group
(i.e. who find themselves buying when most wish to sell, or vice
versa). The essence of the minority game is that each trader
individually wishes to make profit, but that the net effect of
his/her trading
 actions is defined
fully in terms of (or relative to) the actions taken by the other
traders, and that there is a high degree of frustration since it
is impossible for all traders to be successful at the same time.
In spite of its apparent simplicity, the minority game has been
found to exhibit non-trivial behaviour (e.g. phase transitions
separating an ergodic from a highly non-ergodic regime) and to
pose a considerable challenge to the theorist. Its (stochastic)
equations do not obey detailed balance, so there is no equilibrium
state, and understanding the model properly inevitably requires
solving its dynamics.
 An overview of the literature on the MG and its
many variations and extensions can be found in \cite{Chalweb}.

In the original minority game, the information supplied to the
agents upon which to base their trading decisions consisted of the
history of the market.  However, it was realized \cite{Cava99}
that the dynamics of the MG remains largely unaltered if, instead
of the true history of the market, random information is supplied
to the agents; given $\alpha$ (the relative number of possible
values for the external information), the only relevant condition
is that all agents must be given the {\em same} information
(whether sensible or otherwise). This led to a considerable
simplification of theoretical approaches to the MG, since it
reduced the process to a Markovian one. In \cite{SaviManuRiol99}
it was shown that the relevant quantities studied in the minority
game can be scaled such that they become independent of the number
of agents $N$ when this number becomes very large, opening up the
possibility to use statistical mechanical tools.
 An interesting
generalization of the game was the introduction of agents'
decision noise \cite{CavaGarrGiarSher99}, which was shown not only
to improve worse than random behaviour but also, more
surprisingly, to be able to make it better than
random\footnote{Using a phenomenological theory for the
volatility, based on so-called `crowd-anticrowd' cancellations
\cite{Johnson1}, this effect was partially explained in
\cite{Johnson2,Johnson3}.}. The studies \cite{CavaGarrGiarSher99}
and \cite{ChalMars99} finally paved the way for a number of papers
aiming to develop a solvable statistical mechanical theory.

Early theoretical attempts focused on using additive decision
noise to regularize the microscopic stochastic laws and derive
deterministic (non-linear)  continous-time equations, which
minimized a Lyapunov function
\cite{ChalMarsZecc00,MarsChalZecc00}. This approach was remarkably
successful in identifying the location of a phase transition and
in describing correctly some aspects of the behaviour of the
minority game above this phase transition. However, it became
clear later that the deterministic equations were only
approximate, even for $N\to \infty$ (since they neglected relevant
fluctuations). This initiated a debate about how to construct
exact continuous-time microscopic laws for the minority game
\cite{CommTrieste,CommOxford,GarrMoroSher00,MarsChal01a,
MarsChal01b}, in which all participants restricted themselves to
either deterministic or Fokker-Planck equations (some dealing with
both additive and multiplicative decision noise
\cite{GarrMoroSher00}) but without agreeing on the expression to
be used for the diffusion matrix, and all involving different
types of approximations already at the microscopic level. Finally,
in \cite{HeimCool01,CoolHeimSher01} the problems relating to the
continuous-time limit were circumvented by redefining the
equations at the microscopic level directly in terms of full
averages over all possible values of the external information, and
the dynamics of the resulting so-called `batch minority game' was
solved exactly using generating functional methods \cite{DeDo78}
(first without decision noise \cite{HeimCool01}, then also for
general types of decision noise \cite{CoolHeimSher01}).

 In this paper we solve the dynamics of the original (on-line)
 minority game, along the lines of \cite{HeimCool01,CoolHeimSher01} (i.e.  using
 generating functional techniques). We show how the  problems and
 uncertainties
 relating to temporal regularization, which appear to have been responsible for
 generating the debates and  approximations surrounding the proper form
 of the continuous-time microscopic laws, can be solved and resolved in an elegant and
 transparent way by using the (exact) procedure of \cite{BedeLakaShul71} for
 deriving
 a continuous-time master equation without using the
thermodynamic limit. This allows us to write down the full
stochastic microscopic equations (without truncation or
approximation) and solve the model by adapting to the present
continuous-time (on-line) process the generating functional
techniques which were employed successfully for the discrete-time
(batch) process in \cite{HeimCool01,CoolHeimSher01}. The end
result is an exact dynamical macroscopic theory, for general types
of decision noise (additive, multiplicative, etc),  in the form of
closed equations for correlation- and response functions which are
defined via an effective single-trader process, from which the
statics follows as a spin-off. Our equations describe both the
ergodic and the non-ergodic regime of the minority game, including
transients.

The full exact dynamical solution now available also allows us to
make rigorous retrospective statements about the validity or
otherwise of the various approximations proposed in the past, and
to explain under which conditions such approximations could indeed
have led to correct results. More specifically, we (i) show why it
is in principle not allowed to truncate the Kramers-Moyal
expansion of the microscopic process after the Fokker-Planck term
(let alone after the flow term), but why upon doing so one can
still find the correct macroscopic equations (for the present
version of the MG), (ii) confirm that the different diffusion
matrices for the Fokker-Planck term in the process, as proposed
earlier by others, are incomplete or approximate, and (iii)
indicate how previously proposed approximations involving the
market volatility can be traced back to assumptions relating to
ergodicity (and are thus valid at most in the ergodic regime).

\section{Definitions}

\subsection{The Minority Game}

The minority game describes the dynamics of decision making by $N$
interacting trading agents, labeled with Roman indices.  At each
round $\ell$ of the game, each agent $i$ has to take a binary
trading decision $s_i(\ell)\in\{-1,1\}$ (e.g. whether to sell or
buy).  At each round all agents are given the same external
information $I_{\mu(\ell)}$ (representing e.g. the overall state
of the market, political or economic developments, etc.), which
here is chosen randomly and independently from a total number
$p=\alpha N$ of possible values, i.e.
$\mu(\ell)\in\{1,\ldots,\alpha N\}$ for each $\ell$. To generate
trading decisions, each agent $i$ has $S$ fixed strategies
$\bR_{ia}=(R_{ia}^1,\ldots,R_{ia}^{\alpha N})\in\{-1,1\}^{\alpha
N}$ at his/her disposal, with $a\in\{1,\ldots,S\}$.  These
strategies act as manuals (or look-up tables) for decision making:
if strategy $a$ is being used by agent $i$ at stage $\ell$ in the
game, then the observation of external information $\mu(\ell)$
will trigger this particular agent into taking the trading
decision $s_i(\ell)=R_{ia}^{\mu(\ell)}$.  Hence the intrinsic
dynamics of the minority game is not driven by the decision
variables $s_i(\ell)$, but by the dynamic selection by each trader
 of trading strategies from his/her available arsenal, as described
below. Each component $R_{ia}^\mu$ is assumed to have been drawn
randomly and independently from $\{-1,1\}$ before the start of the
game, with uniform probabilities.  The strategies thus introduce
quenched disorder into the game.

Given a choice $\mu(\ell)$ made  at the start of round $\ell$,
every agent $i$ selects the strategy $\tilde{a}_i(\ell)$ which for
trader $i$ has the highest pay-off value at that point in time,
i.e. $\tilde{a}_i(\ell)=\mbox{arg max}~ p_{ia}(\ell)$, and
subsequently makes the binary bid
$b_i(\ell)=R^{\mu(\ell)}_{i\tilde{a}_i(\ell)}$. The (re-scaled)
total bid at stage $\ell$ is defined as $A(\ell)=N^{-1/2}\sum_i
b_i(\ell)$. Next all agents update the pay-off values of {\em
each} of their  strategies $a$ on the basis of what would have
happened if they had played that particular strategy:
\be
  p_{ia}(\ell\!+\!1)=p_{ia}(\ell) -\frac{\rate}{\sqrt{N}} R^{\mu(\ell)}_{ia} A(\ell)
\label{eq:online} \ee The constant $\rate$ represents an
(optional) learning rate. Note that the agents all behave as price
takers, i.e. they do not take into account the impact of their own
decisions on the total bid.
 We here consider only the $S=2$
model, where the equations can be simplified  upon introducing
$q_i(\ell)=\frac{1}{2}[p_{i1}(\ell)-p_{i2}(\ell)]$,
$\bomega_i=\frac{1}{2}[\bR_{i1}+\bR_{i2}]$ and
$\bxi_i=\frac{1}{2}[\bR_{i1}-\bR_{i2}]$. As a result, the selected
strategy in round $\ell$ can now be written as
$\bR_{i\tilde{a}_i(\ell)}=\bomega_i+\sgn[q_i(\ell)]\bxi_i$, and
the evolution of the difference is given by:
\be
  q_i(\ell\!+\!1)=q_i(\ell)-\frac{\rate}{\sqrt{N}} \xi_i^{\mu(\ell)}[
    \Omega^{\mu(\ell)} +
    \frac{1}{\sqrt{N}}\sum_j \xi_j^{\mu(\ell)} \sgn[q_j(\ell)]
    ]
\label{eq:online_S2}
\end{equation}
with $\bOmega=N^{-1/2}\sum_j \bomega_j\in\Re^{\alpha N}$.

\subsection{Minority Game with Decision Noise}

The process (\ref{eq:online_S2}) can and has been generalized in
order to include traders' decision noise
\cite{CavaGarrGiarSher99,ChalMarsZecc00}. This can be done in many
different ways; here we try to avoid being unnecessarily
specific, and choose a general definition where we replace (as in e.g. \cite{CoolHeimSher01})
\begin{equation}
\sgn[q_j(\ell)]~\rightarrow~
\sigma[q_j(\ell),z_j(\ell)]
\label{eq:generalnoise}
\end{equation}
in which the $z_{j}(\ell)$ are independent and
zero average random numbers, described by some symmetric distribution $P(z)$ which is normalised according to
 $\int\!dz~P(z)=\int\!dz~P(z)z^2=1$. The function $\sigma[q,z]$ is
 parametrized by a control parameter $T\geq 0$
 such that $\sigma[q,z]\in \{-1,1\}$, with $\lim_{T\to
 0}\sigma[q,z]=\sgn[q]$ and
 $\lim_{T\to\infty}\int\!dz~P(z)\sigma[q,z]=0$, so that $T$ can be interpreted as a measure
 of the degree of stochasticity in the traders' decision making. Typical examples are
 additive and multiplicative noise definitions such as
\begin{eqnarray}
  {\rm additive:~~~~~~~~~~} && \sigma[q,z]=\sgn[q+Tz]
  \label{eq:additive}
\\
  {\rm multiplicative:~~~} &&
  \sigma[q,z]=\sgn[q]~\sgn[1+Tz]
  \label{eq:multiplicative}
\end{eqnarray}
In the first case (\ref{eq:additive}) the noise has the potential
to be overruled by the so-called `frozen' agents, which have
$q_i(t)\sim \tilde{q}_i t$ for $t\to\infty$ (see
\cite{HeimCool01}).  In the second case the decision noise will
even retain its effect for `frozen' agents (if they exist). These
definitions represent situations where for $T>0$ a trader need not
always use his/her `best' strategy; for $T\to 0$ we revert back to
the process (\ref{eq:online_S2}). The impact of the multiplicative
noise (\ref{eq:multiplicative}) can be characterised by the
monotonic function
\begin{equation}
\lambda(T)=\int\!dz~P(z)~\sgn[1+Tz] \label{eq:lambdaT}
\end{equation}
with $\lambda(0)=1$ and $\lambda(\infty)=0$. For e.g. a Gaussian
$P(z)$ one has $\lambda(T)=\erf[1/\sqrt{2}T]$. We now find
(\ref{eq:online_S2}) being replaced by
\be
  \label{eq:online_S2_noise}
  q_i(\ell\!+\!1)=q_i(\ell)-\frac{\rate}{\sqrt{N}}~ \xi_i^{\mu(\ell)}
  A^{\mu(\ell)}[\bq(\ell),\bz(\ell)],
\end{equation}
In this expression the quantity $A^\mu[\bq,\bz]$ denotes the bid
which would result upon presentation of information $\mu$, given
the microscopic state $\bq$ of the system and given realization
$\bz$ of the decision noise:
\begin{equation} \label{eq:Amu}
    A^\mu[\bq,\bz]=
    \Omega^{\mu} +
    \frac{1}{\sqrt{N}}\sum_j \xi_j^{\mu} \sigma[q_j,z_j]
\end{equation}

\section{Microscopic Probabilistic Description}

\subsection{Temporal Regularization Using the Procedure of Bedeaux et al}

We convert the discrete-time on-line stochastic process
(\ref{eq:online}) into an explicit Markovian description in terms of
probability densities.  For the present noisy version of the game one
finds a microscopic transition probability density operator
$W(\bq|\bq^\prime)$ which involves an average over the random numbers
$\{z_i\}$, indicated by $\bra\ldots\ket_{\bz}$:
\begin{eqnarray}
  p_{\ell+1}(\bq)
&=&
  \int\!d\bq^\prime~W(\bq|\bq^\prime)
  p_\ell(\bq^\prime)
  \label{eq:Markov_chain}
\\
  W(\bq|\bq^\prime)
&=&
  \frac{1}{p}\sum_{\mu=1}^p
  \left\bra \prod_i
  \delta\left[
     q_i-q_i^\prime+\frac{\rate}{\sqrt{N}}
     \xi_i^{\mu} A^\mu[\bq',\bz]
   \right]\right\ket_{\bz}
\label{eq:transition_kernel}
\end{eqnarray}
In order to carry out systematically the temporal coarse-graining
and transform the dynamics to an appropriately re-scaled continuous time $t$,
we follow the systematic procedure of \cite{BedeLakaShul71} and define  the
{\em duration} of each of the iteration steps to be
a continuous random number, the statistics of which are described by the probability $\pi_\ell(t)$
that at time $t$ precisely $\ell$ updates have been
made. Our new process (including the randomness in step duration) is described by
\[
  p_t(\bq)=\sum_{\ell\geq 0}\pi_\ell(t)p_\ell(\bq)
=
  \sum_{\ell\geq 0}\pi_\ell(t)
  \int\!d\bq^{\prime}~W^\ell(\bq|\bq^{\prime})p_0(\bq^{\prime})
\]
and time has become continuous. For $\pi_\ell(t)$ we make the
Poisson choice $\pi_\ell(t)=\frac{1}{\ell!}(t/\Delta_N)^\ell
e^{-t/\Delta_N}$. From $\bra \ell\ket_\pi=t/\Delta_N$ and $\bra
\ell^2\ket_\pi=t/\Delta_N+t^2/\Delta_N^2$ it follows that
$\Delta_N$ is the average duration of an iteration step, and that
the relative deviation in $\ell$ at a given $t$ vanishes for
$\Delta_N\to 0$ as
 $\sqrt{\bra\ell^2\ket_\pi-\bra\ell\ket_\pi^2}/\bra
 \ell\ket_\pi=\sqrt{\Delta_N/t}$.
This introduction of random step durations thus only introduces
uncertainty about where we are on the time axis, which will vanish
at the end of the calculation, provided we ensure $\lim_{N\to
\infty}\Delta_N=0$. The properties of the Poisson distribution
under temporal derivation lead to
\begin{eqnarray}
  \frac{d}{dt}p_t(\bq)
  &=&
  \frac{1}{\Delta_N}\left\{\int\!d\bq^{\prime}~W(\bq|\bq^{\prime})p_t(\bq^{\prime})- p_t(\bq)
\right\} \label{eq:master}
\end{eqnarray}

\subsection{Canonical Temporal Coarse Graining}

 To find the appropriate
scaling with $N$ of $\Delta_N$ (and investigate the relation with
the Fokker-Planck approximations of \cite{GarrMoroSher00} and
\cite{MarsChal01b} in a subsequent section) it is instructive to
expand the master equation (\ref{eq:master}) in powers of the
learning rate:
\begin{eqnarray}
  \frac{d}{dt}p_t(\bq)
&=&
  \frac{1}{\Delta_N p}
  \sum_{\mu=1}^p\int\!d\bq^\prime~p_t(\bq^\prime)
  \bra
    \delta(
      \bq-\bq^\prime
      +\frac{\rate}{\sqrt{N}}\bxi^{\mu}
      A^\mu[\bq',\bz]
    )
    -\delta(\bq-\bq^\prime)
  \ket_{\bz}
\nonumber \\
&=&
\sum_{\ell\geq 1}\left[ L_\ell~ p_t\right](\bq)
\label{eq:KM}
\end{eqnarray}
with
\begin{eqnarray}
  \left[ L_\ell ~p\right](\bq)
&=&
  \frac{\rate^\ell}{N^{\ell/2}}\!\!
  \sum_{n_1,\ldots,n_N\geq 0}\frac{\delta_{\ell,\sum_i n_i}}{n_1!\ldots n_N!}
  \frac{\partial^\ell}{\partial q_1^{n_1}\ldots \partial
  q_N^{n_N}}
  \nonumber \\
&&
  \times\left\{p(\bq)
  \!\!\left[
  \frac{1}{\Delta_N
  p}\!\sum_{\mu=1}^p
  (\xi_1^\mu)^{n_1}\!\!\!\!\!\ldots (\xi_N^\mu)^{n_N}
  \bra
  A^\mu[\bq,\bz]^\ell\ket_{\bz}
  \right]
  \right\}
  \label{eq:KM_terms}
\end{eqnarray}
From the $\ell=1$ term in this so-called Kramers-Moyal expansion
one reads off the canonical scaling for the temporal
coarse-graining time-scale $\Delta_N$ (in order to guarantee  a
proper $N\to\infty$ limit):
\[
  \left[ L_1 ~p\right](\bq)
=
  \frac{\rate}{\alpha\Delta_N N} \sum_i\frac{\partial}{\partial q_i}
  \left\{p(\bq)
  \left[
  \frac{\bxi_i\cdot\bOmega}{\sqrt{N}}
  +\frac{1}{N}\!\sum_j \bxi_i\cdot\bxi_j \bra\sigma[q_j,z]\ket_z
  \right]
  \right\}
\]
We are automatically led to the choice $\Delta_N=\order(\rate/N)$
(this gives, for $T=0$ and for $N\to\infty$, as our first term
exactly the batch dynamics studied in \cite{HeimCool01}), and we
must require $\lim_{N\to\infty}\rate/N=0$ in order to guarantee
$\lim_{N\to\infty}\Delta_N=0$.  We now put
\begin{equation}
\Delta_N=\rate/2\alpha N \label{eq:Delta_found}
\end{equation}
(thereby {\em en passant} absorbing an additional distracting
factor $2\alpha$, to find simpler equations later) and find the
various terms (\ref{eq:KM_terms}) in the Kramers-Moyal expansion
(\ref{eq:KM}) reducing to
\begin{eqnarray}
  \left[ L_\ell ~p\right](\bq)
  &=&
  \frac{2\rate^{\ell-1}}{N^{\frac{1}{2}(\ell-1)}}\!\!
  \sum_{n_1,\ldots,n_N\geq 0}\frac{\delta_{\ell,\sum_i n_i}}{n_1!\ldots n_N!}
  \frac{\partial^\ell}{\partial q_1^{n_1}\ldots \partial
  q_N^{n_N}}
  \nonumber \\
  &&
  \hspace*{-3mm}
  \left\{
    p(\bq)
    \!\!\left[
      \frac{1}{\sqrt{N}}\!\sum_{\mu=1}^p
      (\xi_1^\mu)^{n_1}\!\!\!\!\!\ldots (\xi_N^\mu)^{n_N}
      \bra
        A^\mu[\bq,\bz]^\ell
      \ket_{\bz}
    \right]
  \right\}
  \nonumber\\
  \label{eq:final_KM_terms}
\end{eqnarray}
Note that, according to (\ref{eq:Delta_found}), the present
canonical definition of the time unit $t$ implies temporal
coarse-graining over $\order(1/\Delta_N)=\order(N/\rate)$
iteration steps.

\subsection{Canonical Scaling of the Learning Rate}

The remaining freedom one has is in the choice of the learning
rate $\rate$. In order to find the appropriate choice(s) for the
scaling with $N$ of $\rate$ we work out the temporal derivatives
of the probability density $P_k(q)=\bra\delta[q-q_k]\ket$ of
individual components of the microscopic state vector, where $\bra
f(\bq) \ket=\int\!d\bq~p(\bq)f(\bq)$ (via integration by parts in
the various terms of (\ref{eq:KM})):
\begin{eqnarray}
  \frac{d}{dt}P_k(q)
&=&
  \sum_{\ell\geq 1}
  \frac{2\rate^{\ell-1}}{N^{\ell/2}\ell!}
  \frac{\partial^\ell}{\partial q^\ell}
  \left\bra
    \delta[q-q_k]
    \sum_{\mu=1}^p
    (\xi_k^\mu)^{\ell}\bra
    A^\mu[\bq,\bz]^\ell\ket_{\bz}
  \right\ket
  \nonumber
  \\
&=&
  \frac{\partial}{\partial q} \left\bra
    2\delta[q-q_k] \left[
    \frac{\bxi_k\cdot\bOmega}{\sqrt{N}} +
    \frac{1}{N}\sum_j\bxi_k\cdot\bxi_j \bra\sigma[q_j,z_j]\ket_{\bz}
    \right]
  \right\ket \nonumber\\
&&
  +\rate
  \frac{\partial^2}{\partial q^2} \left\bra
    \delta[q-q_k]
    \frac{1}{N}\sum_{\mu=1}^p \bra
    \left[\xi_k^\mu A^\mu[\bq,\bz] \right]^2\ket_{\bz}
  \right\ket
  \nonumber\\
  && +\!\sum_{\ell>2}\order(N^{1-\frac{1}{2}\ell}\rate^{\ell-1})
  \label{eq:projected_distribution}
\end{eqnarray}
We see explicitly that the diffusion term in this equation is of
order $\order(\rate)$. Hence, in order to prevent the system from
being overruled by fluctuations we have to choose $\rate
=\order(N^{0})$ (with the fluctuations in individual components
vanishing altogether as soon as $\lim_{N\to\infty}\rate=0$).
Following \cite{CavaGarrGiarSher99,ChalMarsZecc00} and subsequent
papers we choose $\rate$ to be a constant which is independent of
$N$ in the remainder of this study. As a consequence we find that
for $N\to\infty$ the single-trader equation
(\ref{eq:projected_distribution}) reduces to a Fokker-Planck
equation, in agreement with \cite{GarrMoroSher00} and
\cite{MarsChal01b} (although, as we will show in a subsequent
section, the diffusion matrices proposed in the latter two studies
are both approximations). This, however, does not necessarily
imply that the underlying $N$-agent process can also be described
by a Fokker-Planck equation.

\section{Properties of the Kramers-Moyal Expansion}

\subsection{Relevance of Higher Orders}

With the definitions $\Delta_N=\rate/2p$ and $\rate=\order(1)$ our
full microscopic master equation (\ref{eq:master}) becomes
\begin{eqnarray}
\hspace*{-9mm} \frac{d}{dt}p_t(\bq)&=&
  \frac{2p}{\rate}\int\!d\bq^{\prime}~p_t(\bq^{\prime})\left\{
    \frac{1}{p}\sum_{\mu=1}^p
  \left\bra
  \delta\left[
     \bq- \bq^\prime\!+ \frac{\rate}{\sqrt{N}}
     \bxi^{\mu} A^\mu[\bq',\bz]
   \right]\right\ket_{\bz}
   \!- \delta(\bq^\prime\!- \bq)\right\}
\nonumber\\ &&
  \label{eq:final_master}
\end{eqnarray}
The Kramers-Moyal expansion  takes the final form
\begin{eqnarray}
  \frac{d}{dt}p_t(\bq)
&=&
  \sum_{\ell\geq 1}\left[ L_\ell~ p_t\right](\bq)
  \label{eq:final_KM}
\end{eqnarray}
where
\begin{eqnarray}
  \left[ L_\ell p\right](\bq)
&=&
  \frac{2\rate^{\ell-1}}{N^{\frac{1}{2}\ell}}
  \sum_{\mu}
  \sum_{n_1,\ldots,n_N\geq 0}\frac{\delta_{\ell,\sum_i n_i}}{n_1!\ldots n_N!}
  (\xi_1^\mu)^{n_1}\!\!\!\!\!\ldots (\xi_N^\mu)^{n_N}
  \frac{\partial^\ell}{\partial q_1^{n_1}\ldots \partial
  q_N^{n_N}}
  \nonumber \\
&&
  \times
  \left\{p(\bq)
  \bra
  A^\mu[\bq,\bz]^\ell\ket_{\bz}
  \right\}
\label{eq:KM_terms_a}
\end{eqnarray}
or, equivalently,
\begin{eqnarray}
  \left[ L_\ell p\right](\bq)
&=&
  \frac{2\rate^{\ell-1}}{\ell !}
  \sum_{\mu}\!
  \left[\frac{1}{\sqrt{N}}\!\sum_i\xi_i^\mu\!\frac{\partial}{\partial
  q_i}\right]^\ell
  \!\left\{
  p(\bq)
  \bra
  A^\mu[\bq,\bz]^\ell\ket_{\bz}
  \right\}
  \label{eq:KM_terms_b}
\end{eqnarray}
Although for $N\to \infty$ the individual components of the state
vector $\bq$ were seen to have Gaussian fluctuations
(\ref{eq:projected_distribution}),
 the cumulative effect  on the dynamics
of the higher order ($\ell>2$) terms in the expansion (\ref{eq:final_KM})
(including cross-correlations) cannot simply
be neglected, since also the {\em number} of components $q_i$ diverges
with $N$. This can be seen more clearly in (\ref{eq:KM_terms_b})
than in (\ref{eq:KM_terms_a}).
If we work out the evolution of averages of observables $f(\bq)$, via
integration by parts, we find that
(\ref{eq:final_KM},\ref{eq:KM_terms_b}) predict
\[
  \rate\frac{d}{dt}\bra f\ket=
  2\sum_{\ell\geq 1}
  \frac{(-\rate)^{\ell}}{\ell !}
  \sum_{\mu}
  \left\bra
    \bra A^\mu[\bq,\bz]^\ell\ket_{\bz}
    \left[\frac{1}{\sqrt{N}}\!\sum_i\xi_i^\mu\!\frac{\partial}{\partial
    q_i}\right]^\ell\! f
  \right\ket
\]
For simple mean-field observables such as $f(\bq)=N^{-1}\sum_i
\zeta_i q_i^{2m}$ (with $m$ independent of $N$) one has
$[\frac{1}{\sqrt{N}}\!\sum_i\xi_i^\mu\!\frac{\partial}{\partial
q_i}]^\ell\! f=\order(N^{-\ell/2})$, and hence the $\ell>2$ terms
in the expansion are of vanishing order. In contrast, for
observables which are dominated by fluctuations, this need no
longer be the case. For instance, in the present system (with its
overall reflection symmetry) one has $f(\bq)
=\frac{1}{\sqrt{N}}\sum_i \zeta_i q_i^{2m+1}=\order(N^0)$; now
$[\frac{1}{\sqrt{N}}\!\sum_i\xi_i^\mu\!\frac{\partial}{\partial
q_i}]^\ell\! f=\order(N^{-(\ell-1)/2})$, and also the $\ell=3$
term could contribute to leading order. Moreover, there is no
practical need to truncate the expansion, since (as we will show
below) within the generating functional formalism it is perfectly
natural to derive an analytical solution of the dynamics with all
terms of the Kramers-Moyal expansion included.

\subsection{Explicit Form of Flow and Diffusion Terms}

Let us now, by way of illustration, work out the first few terms of
(\ref{eq:KM_terms_b}) for additive noise with
$P(z)=\frac{1}{2}K[1-\tanh^2(Kz)]$ (as in
e.g. \cite{ChalMarsZecc00,CavaGarrGiarSher99}, with $K$ such that
$\int\!dz~P(z)z^2=1$), for which $\bra\sigma[q,z]\ket_z=\tanh[\beta
q]$ ($\beta\equiv K/T$):
\begin{eqnarray}
  \left[ L_1 ~p\right](\bq)
&=&
  \sum_i\frac{\partial}{\partial q_i}
  \left\{
   2 p(\bq)
    \left[\frac{\bxi_i\cdot\bOmega}{\sqrt{N}}+\frac{1}{N}\sum_j
    \bxi_i\cdot\bxi_j\tanh[\beta q_j]\right]
  \right\}
  \label{eq:L1_additive}
\\
  \left[ L_2 ~p\right](\bq)
&=&
  \rate
  \sum_{ij}\frac{\partial^2}{\partial q_i\partial q_j}
  \left\{
    p(\bq)\left[
      \frac{1}{N}\!
      \sum_{\mu}
      \xi_i^\mu\xi_j^\mu
      \bra A^\mu[\bq,\bz]^2 \ket_\bz
    \right]
  \right\}
  \label{eq:L2_additive}
\end{eqnarray}
Equation (\ref{eq:L2_additive}) describes, as expected, two types
of fluctuations: $[ L_2 ~p](\bq)= \sum_{k\ell}\partial^2_{k\ell}
\left\{p(\bq)[M_{k\ell}^{A}+M_{k\ell}^B]\right\}$, one describing
fluctuations induced by the randomly and sequentially drawn
external information:
\begin{equation}
  M_{k\ell}^{A}=
\frac{\rate}{N}\!\sum_{\mu=1}^p \xi_k^\mu\xi_\ell^\mu
~(\Omega^\mu+\frac{1}{\sqrt{N}} \sum_j\xi_j^{\mu}\tanh[\beta
q_j])^2 \label{eq:diffusion_A}
\end{equation}
and a second one describing fluctuations induced by the decision
noise\footnote{Note that $\lim_{\beta\to\infty}M_{k\ell}^B=0$, so
that the $\{M_{k\ell}^A\}$ represent the fluctuations due to
external information selection in the absence of decision noise.
For $\beta<\infty$, however, the above causal/interpretational
separation is not perfect, since the two sources of randomness
will inevitably be intertwined.}:
\begin{equation}
  M_{k\ell}^{B}=
  \frac{\rate}{N^2}\sum_{\mu=1}^p
  \xi_k^\mu\xi_\ell^\mu\sum_{i}(\xi_i^\mu)^2(1-\tanh^2[\beta q_i])
  \label{eq:diffusion_B}
\end{equation}
The corresponding expressions for multiplicative noise are
obtained by simply replacing $\tanh[\beta q_k]\to
\lambda(T)\sgn[q_k]$ in equations
(\ref{eq:L1_additive},\ref{eq:diffusion_A},\ref{eq:diffusion_B}).
For both noise types the second contribution
(\ref{eq:diffusion_B}) to the diffusion matrix vanishes in the
limit $\beta\to\infty$ of deterministic decision making.

\section{The Generating Functional}

\subsection{Definition and Discretization}

Instead of working with the Kramers-Moyal expansion, it will be
more convenient and safe to return to the underlying master
equation (\ref{eq:final_master}), which can be written as
\begin{eqnarray}
\frac{d}{dt}p_t(\bq) &=& \frac{2p}{\rate}
  \int\!\frac{d\hat{\bq}d\bq^{\prime}}{(2\pi)^N}
 e^{i\sum_i\hq_i(q_i-q_i^\prime-\theta_i)}  p_t(\bq^{\prime})
\nonumber \\ && \hspace*{10mm} \times
 \left\{
\frac{1}{p}\sum_{\mu=1}^p
  \bra
e^{i\rate (\frac{1}{\sqrt{N}}\sum_i\hq_i\xi_i^{\mu})
A^\mu[\bq',\bz] } \ket_{\bz}
    -1\right\}
\end{eqnarray}
where we have introduced auxiliary driving forces $\theta_i(t)$ in
order to identify response functions later. At this stage we
discretise our continuous time, $t\to \ell \delta$
($\ell=0,1,2,\ldots$) with intervals $0<\delta\ll 1$ which can be
sent to zero {\em independent} of the limit $N\to\infty$ (since
the procedure of \cite{BedeLakaShul71}, as followed here, leads to
a continuous-time master equation for any $N$), so that the
probability density of finding a \emph{path} of microscopic states
$\{\bq(0),\bq(\delta),\bq(2\delta),\ldots\}$ can be written as
\begin{equation}
{\rm Prob}[\bq(0),\bq(\delta),\bq(2\delta),\ldots]=
p_0(\bq(0))\prod_{t>0} \tilde{W}_t[\bq(t)|\bq(t-\delta)]
\end{equation}
with
\begin{eqnarray}
  \tilde{W}_t[\bq|\bq^\prime]
&=&
  \int\!\frac{d\hat{\bq}}{(2\pi)^N}
  e^{i\sum_i\hq_i(q_i-q_i^\prime-\theta_i(t))}
  \nonumber\\
&&
  \times\left\{
  \frac{2\delta}{\rate}
  \sum_\mu \bra
  e^{i\rate[\frac{1}{\sqrt{N}} \sum_i\hq_i\xi_i^{\mu}]
  A^\mu[\bq',\bz]} \ket_{\bz}
  +(1-\frac{2p\delta}{\rate})
\right\}
\end{eqnarray}
The moment generating functional for our stochastic process is defined as
\begin{eqnarray*}
  Z[\bpsi]
  &=&
  \langle~e^{i\sum_t \sum_i  \psi_i(t) q_i(t)}~\rangle
  \\
  &=&
  \int\prod_t\left\{ d\bq(t)~ \tilde{W}_t[\bq(t+\delta)|\bq(t)]\right\}~p_0(\bq(0))~e^{i\sum_t \sum_i  \psi_i(t) q_i(t)}
\end{eqnarray*}
where, as in \cite{HeimCool01}, we introduced external forces
$\theta_i(t)$ to generate response functions later. Derivation of
the generating functional with respect to the conjugate variables
$\bpsi$ generates all moments of $\bq$ at arbitrary times. Upon
introducing the two short-hands:
\[
  w^\mu_t=\frac{\sqrt{2}}{\sqrt{N}}\sum_i \hq_i(t) \xi^\mu_i, \qquad
  x^\mu_t=\frac{\sqrt{2}}{
  \sqrt{N}}\sum_i s_i(t)\xi^\mu_i,
\]
with $s_i(t)\equiv\sigma[q_i(t),z_i(t)]$, as well as
$\D\bq=\prod_{it}[dq_i(t)/\sqrt{2\pi}]$, $\D\bw=\prod_{\mu
t}[dw^\mu_t/\sqrt{2\pi}]$ and $\D\bx=\prod_{\mu
t}[dx^\mu_t/\sqrt{2\pi}]$ (with similar definitions for
$\D\hat{\bq}$, $\D\hat{\bw}$ and $\D\hat{\bx}$, respectively), the
generating functional takes the following form:
\begin{eqnarray}
  Z[\bpsi]
  &=&
   \!\int\!\! \D\bw \D\hat{\bw} \D\bx \D\hat{\bx}~
  e^{i\sum_{t \mu}[\hw^\mu_t w^\mu_t +
        \hx^\mu_t x^\mu_t]}
 \prod_t\!
 \left\{
 1\!+\!
\frac{2\delta}{\rate} \sum_\mu\left[ e^{\frac{i\rate
w_t^\mu}{\sqrt{2}}[\Omega^{\mu}+\frac{x_t^\mu}{\sqrt{2}}]}\!-\!1
\right] \right\}
   \nonumber\\
   &&
  \times \int\! \D\bq \D\hat{\bq}~p_0(\bq(0))~
  \bra
e^{-\frac{i\sqrt{2}}{\sqrt{N}}\sum_{\mu i}\xi_i^\mu\sum_t
[\hat{w}_t^\mu\hq_i(t)
  +\hat{x}^\mu_t s_i(t)]}
  \ket_{\bz}
  \nonumber\\
  &&
 \times
e^{i\sum_{t i}\left[ \hq_i(t)(q_i(t+\delta)-q_i(t)-\theta_i(t)) +
\psi_i(t)q_i(t) \right]}
 \label{eq:Zbeforeaverage1}
\end{eqnarray}
We note that, as $\delta\to 0$ (and using $\delta\sum_t
=\order(\delta^0)$):
\[
  \prod_t\!
 \left\{
 1\!+\!
\frac{2\delta}{\rate} \sum_\mu\left[ e^{\frac{i\rate
w_t^\mu}{\sqrt{2}}[\Omega^{\mu}+\frac{x_t^\mu}{\sqrt{2}}]}\!-\!1
\right] \right\}
=
e^{\sum_\mu \sum_t \frac{2\delta}{\rate}\left[e^{\frac{i\rate
w_t^\mu}{\sqrt{2}}[\Omega^{\mu}+\frac{x_t^\mu}{\sqrt{2}}]}-1
+\order(\delta N)\right] }
\]
Hence, if we choose $\delta\ll N^{-1}$ we obtain
\begin{eqnarray}
  Z[\bpsi]
&=&
  \!\int\!\! \D\bw \D\hat{\bw} \D\bx \D\hat{\bx}~
  e^{\sum_{t \mu}\left[
    i\hw^\mu_t w^\mu_t +
    i\hx^\mu_t x^\mu_t
    + \frac{2\delta}{\rate}[ e^{i\rate w_t^\mu
    [\Omega^{\mu}+\frac{x_t^\mu}{\sqrt{2}}]/\sqrt{2}}-1 +{\smallo}(N^0)]
  \right]} \nonumber\\
&&
  \times \int\! \D\bq \D\hat{\bq}~p_0(\bq(0))~
  \bra
    e^{-\frac{i\sqrt{2}}{\sqrt{N}}\sum_{\mu i}\xi_i^\mu\sum_t
    [\hat{w}_t^\mu\hq_i(t)
    +\hat{x}^\mu_t s_i(t)]}
  \ket_{\bz}
  \nonumber\\
&&
  \times e^{i\sum_{t i}\left[
    \hq_i(t)(q_i(t+\delta)-q_i(t)-\theta_i(t)) + \psi_i(t)q_i(t)
  \right]}
  \label{eq:Zbeforeaverage2}
\end{eqnarray}

\subsection{Disorder Average}

 At this stage we carry
out the disorder averages, denoted as $[\cdots]_{dis}$, which
involve the variables
$\xi_i^\mu=\frac{1}{2}(R^\mu_{i1}-R^\mu_{i2})$ and
$\Omega^\mu=\frac{1}{2}N^{-\frac{1}{2}}\sum_j(R^\mu_{j1}+R^\mu_{j2})$.
For times which do not scale with $N$ and for simple initial
conditions of the form $p_0(\bq)=\prod_i p_{0}(q_i)$ one  finds:
\begin{eqnarray}
 \left[ Z[\bpsi] \right]_{dis}
  &=&
   \!\int\!\! \D\bw \D\hat{\bw} \D\bx \D\hat{\bx}~
  e^{i\sum_{t \mu}[\hw^\mu_t w^\mu_t +
        \hx^\mu_t x^\mu_t]}
   \nonumber \\
    &&
  \hspace*{-15mm}\times \int\! \D\bq \D\hat{\bq}~\prod_i p_0(q_i(0))~
  e^{i\sum_{t i}\left[
    \hq_i(t)(q_i(t+\delta)-q_i(t)-\theta_i(t)) +
    \psi_i(t)q_i(t)
  \right]}
  \label{eq:Zwithaverage1}\\
&&
  \hspace*{-15mm}\times\left[
    \left\bra \prod_\mu
    e^{
      \frac{2\delta}{\rate}\sum_{t}\left[ e^{i\rate w_t^\mu
      [\Omega^\mu+\frac{x_t^\mu}{\sqrt{2}}]/\sqrt{2}}-1 +{\smallo}(N^0)\right]
      -\frac{i\sqrt{2}}{\sqrt{N}}\sum_{i}\xi_i\sum_t
      [\hat{w}_t^\mu\hq_i(t)
      +\hat{x}^\mu_t s_i(t)]
    }
    \right\ket_{\bz}
  \right]_{dis}\nonumber
\end{eqnarray}
We concentrate on the term in (\ref{eq:Zwithaverage1}) with the
disorder averages.
 These are similar to but generally more
difficult than those calculated in
 \cite{HeimCool01} (to which they reduce only for $\rate\to 0$). As in \cite{HeimCool01}
 they automatically generate
 the dynamic order parameters $C_{t
t^\prime}=N^{-1}\sum_i s_i(t) s_i(t^\prime)$, $K_{t
t^\prime}=N^{-1}\sum_i s_i(t) \hq_i(t^\prime)$, and $L_{t
t^\prime}=N^{-1}\sum_i \hq_i(t) \hq_i(t^\prime)$ and their
conjugates:
\begin{eqnarray}
  \prod_\mu \left[\ldots\right]_{dis}
&=&
  \prod_\mu\left\{
    \int\!\frac{d\Omega d\hOmega}{2\pi}~e^{i\hOmega\Omega +
    \frac{2\delta}{\rate}\sum_{t}[ e^{i\rate w_t^\mu
    [\Omega+\frac{x_t^\mu}{\sqrt{2}}]/\sqrt{2}}-1 +\smallo(N^0)]}
  \right. \nonumber\\
&&
  \left.
    \hspace*{-20mm}
    \times
    \prod_i \left[
      e^{ -\frac{iR_{i1}^\mu}{\sqrt{2N}} \left[
        \hOmega/\sqrt{2} +\sum_t [\hat{w}_t^\mu\hq_i(t)
        +\hat{x}^\mu_t s_i(t)]
      \right]
        -\frac{iR_{i2}^\mu}{\sqrt{2N}}\left[
        \hOmega/\sqrt{2} -
        \sum_t[\hat{w}_t^\mu\hq_i(t)
        +\hat{x}^\mu_t s_i(t)]
      \right]}
    \right]_{dis}
  \right\} \nonumber\\
&=&
  \prod_\mu\left\{
\int\!\frac{d\Omega d\hOmega}{2\pi}~e^{i\hOmega\Omega +
\frac{2\delta}{\rate}\sum_{t}[ e^{i\rate w_t^\mu
[\Omega+\frac{x_t^\mu}{\sqrt{2}}]/\sqrt{2}}-1] +\smallo(N^0)}
\right.
 \nonumber\\
 &&
 \left.
 \times ~ e^{ -\frac{1}{4N}\sum_i \left[
 (\frac{\hOmega}{\sqrt{2}}
+\sum_t [\hat{w}_t^\mu\hq_i(t)
  +\hat{x}^\mu_t s_i(t)])^2
+(\frac{\hOmega}{\sqrt{2}} - \sum_t[\hat{w}_t^\mu\hq_i(t)
  +\hat{x}^\mu_t s_i(t)])^2\right]}
 \right\}
\nonumber\\
 &=&
  \prod_\mu\left\{
\int\!\frac{d\Omega
d\hOmega}{2\pi}~e^{i\hOmega\Omega-\frac{1}{4}\hOmega^2 +
\frac{2\delta}{\rate}\sum_{t}[ e^{i\rate w_t^\mu
[\Omega+\frac{x_t^\mu}{\sqrt{2}}]/\sqrt{2}}-1] +\smallo(N^0)}
\right.
 \nonumber\\
 &&
 \left.
 \times ~ e^{ -\frac{1}{2N}\sum_i \left[
\sum_t [\hat{w}_t^\mu\hq_i(t)
  +\hat{x}^\mu_t s_i(t)]\right]^2}
 \right\}
 \nonumber\\
 &=&
  \prod_\mu\left\{
  \int\!\frac{d\Omega}{\sqrt{\pi}}~ e^{-\Omega^2
  + \frac{2\delta}{\rate}\sum_{t}[
  e^{i\rate w_t^\mu [\Omega+\frac{x_t^\mu}{\sqrt{2}}]/\sqrt{2}}-1]
  +\smallo(N^0)} \right.
  \nonumber\\
&&
  \left.
  \hspace*{20mm}
  \times~ e^{
  -\frac{1}{2}\sum_{tt^\prime}
  \left[
    \hat{w}_t^\mu L_{t t^\prime} \hat{w}_{t^\prime}^\mu
    +2\hat{x}^\mu_t K_{t t^\prime} \hat{w}_{t^\prime}^\mu
   +\hat{x}^\mu_t C_{t t^\prime} \hat{x}^\mu_{t^\prime}
  \right]} \right\}
\label{eq:disorder}
\end{eqnarray}
Insertion into (\ref{eq:Zwithaverage1}), followed by the isolation
of the order parameters via $\delta$-distributions (whose integral
representations generate the conjugate order parameters) then
gives
\begin{equation}
  \overline{Z[\bpsi]}=
  \int\![\D C \D\hat{C}][\D K \D\hat{K}][ \D L \D\hat{L}]~
    e^{N\left[\Psi+\Phi+\Omega +\smallo(N^0)\right]}
    \label{eq:Zafteraverage}
\end{equation}
The $\smallo(N^0)$ term in the exponent is independent of the fields
$\{\psi_i(t)\}$ and $\{\theta_i(t)\}$. Upon choosing an
appropriate scaling with $\delta$ of the conjugate integration
variables  (viz. $(\hat{x},\hat{w})\to \delta (\hat{x},\hat{w})$
and $(\hat{C},\hat{K},\hat{L})\to \delta^2
(\hat{C},\hat{K},\hat{L})$, to guarantee the existence of a proper
$\delta\to 0$ limit of the various time summations), the three
relevant exponents in (\ref{eq:Zafteraverage}) are given by the
following expressions:
\begin{eqnarray}
\Psi&=&i\delta^2\sum_{tt^\prime}[ \hC_{tt^\prime} C_{tt^\prime}
+\hK_{t t^\prime} K_{tt^\prime} + \hL_{tt^\prime} L_{tt^\prime}],
\label{eq:Psi}
\\
\Phi &=& \alpha  \log\left[ \int\! \D w \D\hat{w} \D x \D\hat{x}~
  e^{
 -\frac{1}{2}\delta^2\sum_{tt^\prime}
 \left[
\hat{w}_t L_{t t^\prime} \hat{w}_{t^\prime}+2\hat{x}_t K_{t
t^\prime} \hat{w}_{t^\prime} +\hat{x}_t C_{t t^\prime}
\hat{x}_{t^\prime} \right]
        }
        \right.
   \nonumber\\
    &&
    \left.
\times ~ e^{i\delta\sum_{t}[\hw_t w_t +
        \hx_t x_t]} \int\!Du~
        e^{\frac{2\delta}{\rate}\sum_{t}[ e^{\frac{1}{2}i\rate
w_t[u+x_t]}-1]} \right]
 \label{eq:Phi}
\\
\Omega &=& \frac{1}{N}\!\sum_i\log
 \bra \int\! \D q \D\hat{q}~p_0(q(0))~
  e^{i\delta\sum_{t}\left[ \hq(t)(\frac{q(t+\delta)-q(t)}{\delta}-\delta^{-1}\theta_i(t)) +
\delta^{-1}\psi_i(t)q(t) \right] }
 \nonumber\\ &&
\times~ e^{-i\delta^2\sum_{tt^\prime}[
\hC_{tt^\prime}s(t)s(t^\prime) +\hK_{t t^\prime}
 s(t)\hq(t^\prime)
+ \hL_{tt^\prime}\hq(t)\hq(t^\prime)]}
  \ket_{\bz}
\label{eq:Omega}
\end{eqnarray}
with the standard abbreviation of the Gaussian measure
$Du=(2\pi)^{-\frac{1}{2}}e^{-\frac{1}{2}u^2}du$.
 The average $\bra \ldots\ket_{\bz}$ has now been reduced
to a single site one: $\bra g[z_1,z_2,\ldots]\ket_{\bz}
=\int\!\prod_t [dz_t P(z_t)]~g[z_1,z_2,\ldots]$. Following
\cite{HeimCool01} we have also introduced the short-hands $\D
q=\prod_{t}[dq(t)/\sqrt{2\pi}]$, $\D
w=\prod_{t}[dw_t/\sqrt{2\pi}]$, $\D x=\prod_{t}[dx_t/\sqrt{2\pi}]$
(with similar definitions for $\D\hat{q}$, $\D\hat{w}$ and
$\D\hat{x}$).

\section{The Saddle-Point Equations}

\subsection{Derivation of Saddle-Point Equations}

We can now evaluate (\ref{eq:Zafteraverage}) by saddle-point
integration, in the limit $N\to\infty$, and provided the order
parameter functions depend asymptotically on their two time
arguments in a sufficiently smooth (i.e. $N$-independent) way. We
define $G_{t t^\prime}=-iK_{t t^\prime}$. Taking derivatives with
respect to the generating fields
 and using the normalisation $\overline{Z[{\bf
0}]}=1$ then gives (at the physical saddle-point) the usual relations
\begin{eqnarray}
C_{t t^\prime}&=&\lim_{N\to\infty}\frac{1}{N}\sum_i\overline{\bra
s_i(t) s_i(t^\prime)\ket}
\label{eq:meaningof_C}
\\
G_{t
t^\prime}&=&\lim_{N\to\infty}\frac{1}{N}\sum_i\frac{\partial}{\partial
\theta_i(t^\prime)}\overline{\bra s_i(t)\ket}
\label{eq:meaningof_K}
\\
L_{t t^\prime}&=&0
\label{eq:meaningof_L}
\end{eqnarray}
 Putting
$\psi_i(t)=0$ (they are no longer needed) and
$\theta_i(t)=\delta.\tilde{\theta}(t)$ then simplifies
(\ref{eq:Omega}) to
\begin{eqnarray}
  \Omega
&=&
  \log
  \int\! \D q \D\hat{q}~p_0(q(0))~
  e^{-i\delta^2\sum_{tt^\prime}\hq(t)\hL_{t t^\prime}\hq(t^\prime)}
  \nonumber\\
&&
  \times \bra e^{i\delta\sum_{t}
  \hq(t)[\frac{q(t+\delta)-q(t)}{\delta}-\tilde{\theta}(t)
  -\delta\sum_{t^\prime}\hK_{t^\prime t}s(t^\prime)]
  -i\delta^2\sum_{tt^\prime}s(t)\hC_{tt^\prime}s(t^\prime)
}
  \ket_{\bz}
\label{eq:Omega_new}
\end{eqnarray}
in which now $s(t)=\sigma[q(t),z_t]$. Extremisation of the
extensive exponent $\Psi+\Phi+\Omega$ of (\ref{eq:Zafteraverage})
with respect to $\{C,\hat{C},K,\hat{K},L,\hat{L}\}$ gives the
remaining saddle-point equations
\begin{eqnarray}
C_{t t^\prime}=\langle s(t) s(t^\prime) \rangle_\star ~~~~~~~~
G_{t t^\prime}=\frac{\partial \bra
s(t)\ket_\star}{\delta\partial\tilde{\theta}({t^\prime})} ~~~~~~~~
\label{eq:CandG}
\\
\hC_{tt^\prime}=\frac{i}{\delta^2}\frac{\partial \Phi}{\partial
C_{t t^\prime}} ~~~~~~~~~~
\hK_{tt^\prime}=\frac{i}{\delta^2}\frac{\partial \Phi}{\partial
K_{t t^\prime}} ~~~~~~~~~~
\hL_{tt^\prime}=\frac{i}{\delta^2}\frac{\partial \Phi}{\partial
L_{t t^\prime}} \label{eq:Conjugates}
\end{eqnarray}
The effective single-trader
averages $\bra \ldots\ket_\star$, generated by taking
derivatives of (\ref{eq:Omega}), are defined as
\begin{equation}
\bra f[\{q,s\}]\ket_\star= \frac{\int\! \D q ~\bra
M[\{q,s\}]f[\{q,s\}]\ket_{\bz}}{\int\! \D q ~\bra
M[\{q,s\}]\ket_{\bz}} \label{eq:effective_measure}
\end{equation}
\vspace*{-3mm}
\begin{eqnarray}
M[\{q,s\}]&=& p_0(q(0))~e^{-i\delta^2\sum_{t t^\prime} s(t)
\hat{C}_{t t^\prime}s(t^\prime)} \int\!
\D\hat{q}~e^{-i\delta^2\sum_{t t^\prime}\hq(t)\hat{L}_{t
t^\prime}\hq(t^\prime)} \nonumber \\ && \times~ e^{i\delta\sum_{t}
  \hq(t)[\frac{q(t+\delta)-q(t)}{\delta}-\tilde{\theta}(t)
  -\delta\sum_{t^\prime}\hK_{t^\prime t}s(t^\prime)]
} \label{eq:singletrader_measure}
\end{eqnarray}
Upon elimination of the trio $\{\hat{C},\hat{K},\hat{L}\}$ via
(\ref{eq:Conjugates}) we obtain exact closed equations for the
disorder-averaged correlation- and response functions in the
$N\to\infty$ limit: equations (\ref{eq:CandG}), with the effective
single trader measure (\ref{eq:singletrader_measure}).

\subsection{Simplification of Saddle-Point Equations}

The introduction of on-line evolution and decision noise into the
dynamics has affected the terms $\Phi$ (\ref{eq:Phi}) and $\Omega$
(\ref{eq:Omega_new}), compared to the analysis in
\cite{HeimCool01}. In order to proceed we now have to work out the
term $\Phi$ (\ref{eq:Phi}) further:
\begin{eqnarray}
  \Phi
&=&
  \alpha  \log  \int\!  Du \D w \D x~\phi[\{x,w\},u] \nonumber
  \\
&&
  \times \int\! \D\hat{w} \D\hat{x}~
  e^{ -\frac{1}{2}\delta^2\sum_{tt^\prime}
  \left[
    \hat{w}_t L_{t t^\prime} \hat{w}_{t^\prime}
    +2\hat{x}_t K_{t t^\prime} \hat{w}_{t^\prime}
    +\hat{x}_t C_{t t^\prime}\hat{x}_{t^\prime}
  \right]
  +i\delta\sum_{t}[\hw_t w_t +\hx_t x_t]}
  \label{eq:Phi_workingout1}
\end{eqnarray}
where
\begin{equation}
  \phi[\{x,w\},u]
=
  \exp\left[
    \frac{2\delta}{\rate}\sum_{t}[ e^{\frac{1}{2}i\rate
    w_t[u+x_t]}-1]
  \right]
\end{equation}
Upon expanding the `inner' exponential in this expression in a
power series, we can write $\phi$ as an average of the form:
\begin{equation}
  \phi[\{x,w\},u]
=
  \left\bra
    \exp\left[ \frac{1}{2}i\rate \sum_t n_t w_t[u+x_t]\right]
  \right\ket_n
  \label{eq:Poissonterm}
\end{equation}
where $\bra \cdot \ket_n$ is defined as:
\begin{equation}
  \bra f[n_1,n_2,\ldots] \ket_{n}
\equiv
  \sum_{n_1,n_2\ldots \geq 0}
  \left[
    \prod_s P_{2\delta/\rate}[n_s]~f[n_1,n_2,\ldots]
  \right]
\end{equation}
with  the Poisson distribution $P_a[\ell]=\frac{1}{\ell
!}e^{-a}a^\ell$. The first two moments of the distribution are
given by $\bra n_t \ket_n=2\delta/\rate$ and $\bra n_t^2
\ket_n=(2\delta/\rate)^2 + 2\delta/\rate$.
  We now insert
(\ref{eq:Poissonterm}) into (\ref{eq:Phi_workingout1}), followed
by integration over $u$, $\{x\}$, $\{\hat{x}\}$ and $\{w\}$ (in
precisely that order). This gives, with $K=iG$, with the diagonal
matrix $E_{tt^\prime}=\frac{1}{2}\rate n_t\delta_{tt^\prime}$, and
with the matrix $D$ whose entries are defined as
$D_{tt^\prime}=1+C_{tt^\prime}$:
\begin{eqnarray}
  \Phi
&=&
  \alpha\log
    \left\bra
      {\rm Det}^{-\frac{1}{2}}[EDE]
      \right.\nonumber\\
      &&\left.\times
      \int\!\D \hat{w}~e^{-\frac{1}{2}\delta^2
      \sum_{tt^\prime}\hat{w}_t\left[
        L+(\openone+ G^\dag E)(EDE)^{-1}(\openone+ EG)
      \right]_{tt^\prime}\hat{w}_{t^\prime} }
    \right\ket_n
  \label{eq:Phi_workingout2}
\end{eqnarray}
(modulo an irrelevant constant). Taking the derivative of $\Phi$
with respect to the matrix elements
$\{L_{tt^\prime},C_{tt^\prime},G_{tt^\prime}\}$, followed by
setting $L\to 0$, gives (using the causality property
$G_{tt^\prime}=0$ for $t\leq t^\prime$, which guarantees that
${\rm Det}(\openone+ EG)=1$):
\begin{eqnarray}
  \lim_{L\to 0}\frac{\partial \Phi}{\partial L_{t t^\prime}}
&=&
  -\frac{1}{2}\alpha \left\bra
    \left[(\openone\!+\! EG)^{-1}EDE(\openone\!+\!G^\dag\!
    E)^{-1}\right]_{t t^\prime}
  \right\ket_n \label{eq:Phi_Lderivative}
\\
  \lim_{L\to 0}\frac{\partial\Phi}{\partial C_{tt^\prime}}
&=&
  0
  \label{eq:Phi_Cderivative}
\\
  \lim_{L\to 0}\frac{\partial\Phi}{\partial G_{tt^\prime}}
&=&
  \!
  -\alpha\left\bra
    \left[(\openone\!+\! EG)^{-1}E\right]_{t^\prime t}
  \right\ket_n
  \label{eq:Phi_Gderivative}
\end{eqnarray}
According to (\ref{eq:Conjugates}) this gives for the conjugate
order parameters:
\begin{eqnarray}
  \hL_{tt^\prime}&=&
  -\frac{1}{2}i\alpha \Sigma_{tt^\prime}
  \label{eq:found_Lhat}
\\
  \hC_{tt^\prime}
&=&
  0
  \label{eq:found_Chat}
\\
  \hK_{tt^\prime}
&=&
  -\alpha R_{t^\prime t}
  \label{eq:found_Khat}
\end{eqnarray}
with
\begin{eqnarray}
  \Sigma
&=&
  \frac{1}{\delta^2}\left\bra
    (\openone\!+\! EG)^{-1}EDE(\openone\!+\! G^\dag\! E)^{-1}
  \right\ket_n \label{eq:noise_covariance}
  \\
  R
&=&
  \frac{1}{\delta^2}\left\bra
    (\openone\!+\! EG)^{-1} E
  \right\ket_n \label{eq:defineR}
\end{eqnarray}

\subsection{Evaluation of Poisson Averages}

Finally, it turns out that in these latter two matrices
(\ref{eq:noise_covariance},\ref{eq:defineR}) the averages over the
$\{n_t\}$
 can be performed exactly, by using causality.
We first deal with (\ref{eq:defineR}) (which is simpler):
\[
R_{t t^\prime}= \frac{1}{\delta^2}\sum_{\ell\geq 0}(-1)^\ell \bra
[(EG)^\ell E]_{tt^\prime}\ket_n
\]
We note that each of the terms in the expansion gives factorised
averages due to $G_{ss^\prime}=0$ for $s\leq s^\prime$ (using only
that $\bra n_s\ket_n=2\delta/\rate$ for any $s$):
\begin{eqnarray}
  \bra [(EG)^\ell E]_{tt^\prime}\ket_n&=&
  (\frac{1}{2}\rate)^{\ell+1}\!\!\!\!\sum_{s_1,\ldots,s_{\ell-1}}\!\!\! \bra n_t
  G_{t s_1} n_{s_1} G_{s_1 s_2} n_{s_2} G_{s_2 s_3} \ldots
  n_{s_{\ell-1}} G_{s_{\ell-1} t^\prime} n_{t^\prime}\ket_n
  \nonumber
  \\
&=& (\frac{1}{2}\rate)^{\ell+1}\!\!\!\!\!\!\!
   \sum_{t>s_1>\ldots >s_{\ell-1}>t^\prime}\!\!\! \!\!\!
   G_{t s_1} G_{s_1 s_2} \ldots  G_{s_{\ell-1} t^\prime}
  \bra n_t n_{s_1}
  \ldots n_{s_{\ell-1}} n_{t^\prime}\ket_n \nonumber
  \\
&=&
  \delta^{\ell+1} G^\ell_{t t^\prime} \nonumber
\end{eqnarray}
giving the simple result
\begin{eqnarray}
  R&=& \frac{1}{\delta}[\openone+\delta G]^{-1}
  \label{eq:finishedR}
\end{eqnarray}
We note that in calculating (\ref{eq:finishedR}) knowledge of only
the first moment of the $\{n_t\}$ distribution was required.

Next we turn to the noise covariance matrix:
\begin{equation}
\Sigma_{t t^\prime}=\frac{1}{\delta^2}\sum_{\ell \ell^\prime\geq
0}(-1)^{\ell+\ell^\prime}
 \bra [(EG)^\ell EDE (G^\dag\! E)^{\ell^{\prime}}]_{t t^\prime} \ket_n
 \label{eq:final_sigma}
\end{equation}
Now, again due to causality, only averages of terms with at most
{\em two} $n$-variables with the same time index can occur. We
have to take into account the extra contributions coming from all
possible time pairings.  Using $\bra n_s^2\ket=\bra
n_s\ket^2[1+\rate/2\delta]$ we derive:
\begin{eqnarray*}
  \bra [(EG)^\ell EDE (G^\dag\! E)^{\ell^{\prime}}]_{s_0 s_0^\prime} \ket_n
&=&
  \sum_{s_\ell s_\ell^\prime}
  D_{s_\ell s_\ell^\prime}
  \bra [(EG)^\ell E]_{s_0 s_\ell} [(EG)^{\ell^\prime} E]_{s_0^\prime s_\ell^\prime}
  \ket_n
\end{eqnarray*}
\vspace*{-4mm}
\begin{eqnarray}
  ~~~~~~~~~
&=&
  (\frac{1}{2}\rate)^{\ell+\ell^{\prime}+2}
  \sum_{s_0>\ldots >s_\ell}
  \sum_{s_0^\prime >\ldots >s_\ell^\prime}
  D_{s_\ell s_\ell^\prime}~
  G_{s_0 s_1} \ldots  G_{s_{\ell-1} s_\ell}
  ~ G_{s_0^\prime s^\prime_1} \ldots  G_{s^\prime_{\ell^\prime-1}
  s_\ell^\prime} \nonumber \\
&&
  \hspace*{10mm}
  \times
  \bra n_{s_0}\ldots  n_{s_\ell}
  n_{s_0^\prime} \ldots  n_{s_\ell^\prime}\ket_n
  \nonumber \\
&=&
  \delta^{\ell+\ell^{\prime}+2}
  \sum_{s_0>\ldots >s_\ell}
  \sum_{s_0^\prime >\ldots >s_\ell^\prime}
   D_{s_\ell s_\ell^\prime}
  \left[1+\frac{\rate}{2\delta}
  \right]^{\sum_{i=0}^{\ell}\sum_{j=0}^{\ell^\prime}\delta_{s_i s_j^\prime}}
  \nonumber \\
&&
  \hspace*{10mm}\times
  ~ G_{s_0 s_1} \ldots  G_{s_{\ell-1} s_\ell}
  ~ G_{s_0^\prime s^\prime_1} \ldots  G_{s^\prime_{\ell^\prime-1}
  s_\ell^\prime} \nonumber \\
&=&
  \delta^{\ell+\ell^{\prime}+2}
  \sum_{s_0>\ldots >s_\ell}
  \sum_{s_0^\prime >\ldots >s_\ell^\prime}
  D_{s_\ell s_\ell^\prime}~
  \prod_{i=0}^{\ell} \prod_{j=0}^{\ell^\prime}\left[1+\delta_{s_i
  s^\prime_j}\frac{\rate}{2\delta}\right] \nonumber \\
&&
  \hspace*{10mm}\times
  ~ G_{s_0 s_1} \ldots  G_{s_{\ell-1} s_\ell}
  ~ G_{s_0^\prime s^\prime_1} \ldots  G_{s^\prime_{\ell^\prime-1}
  s_\ell^\prime}
  \label{eq:nasty_beast}
\end{eqnarray}
where the factor
$\sum_{i=0}^{\ell}\sum_{j=0}^{\ell^\prime}\delta_{s_i s_j^\prime}$
counts the number of pairings  occurring in the two types of time
arguments (i.e. those with primes, and those without). We note
that to find expression (\ref{eq:nasty_beast}) knowledge of only
the first two moments of the $\{n_t\}$-distribution was required.

\section{The Effective Single-Agent Process}

\subsection{Discretized Single-Agent Process}

Upon inserting the results
(\ref{eq:found_Lhat},\ref{eq:found_Chat},\ref{eq:found_Khat}) we
now find our effective single trader measure $M[\{q,s\}]$ of
(\ref{eq:singletrader_measure}) reducing further to the following
expression (modulo a constant pre-factor reflecting normalisation,
which is independent of the decision noise variables $\{z(t)\}$):
\begin{eqnarray}
  M[\{q,s\}]&=& p_0(q(0))\int\! \D\eta~e^{-\frac{1}{2}\sum_{t
  t^\prime}\eta(t)\Sigma^{-1}_{t t^\prime}\eta(t^\prime)}\nonumber
\\&&
  \times
  \delta\left[
    \frac{q(t+\delta)-q(t)}{\delta}-\tilde{\theta}(t)
    +\alpha\delta\sum_{t^\prime} R_{t
    t^\prime}s(t^\prime)-\sqrt{\alpha}\eta(t)
  \right]
\label{eq:singleagent_statistics}
\end{eqnarray}
with $R_{tt^\prime}=\delta^{-1}[\openone+\delta G]^{-1}_{t
t^\prime}$ (\ref{eq:finishedR}).
 This describes a  single-agent process of the form
\begin{eqnarray}
\frac{q(t+\delta)-q(t)}{\delta} &=&
\tilde{\theta}(t)
  -\alpha\delta\sum_{t^\prime} R_{t
  t^\prime} \sigma[q(t^\prime),z(t^\prime)]+\sqrt{\alpha}\eta(t)
\label{eq:singleagent}
\end{eqnarray}
Causality ensures that $R_{tt^\prime}=0$ for $t^\prime
>t$. The variable $z_t$ represents the original single-trader
decision noise, with $\bra z(t)\ket_z=0$ and $\bra z(t)
z(t^\prime)\ket_z=\delta_{tt^\prime}$, and
 $\eta(t)$ is a disorder-generated Gaussian noise with zero mean and with
  temporal correlations given by (\ref{eq:final_sigma},\ref{eq:nasty_beast}).

\subsection{Continuous Time Limit}

 We can now take the limit $\delta\to 0$ and restore
continuous time. The bookkeeping of $\delta$-terms is found to
come out right, with partial derivatives with respect to the
perturbation fields converting into functional derivatives, with
time summations converting into integrals, and with matrices
converting into integral operators (with the usual convention
$\frac{1}{\delta} \openone_{tt^\prime}\to \delta[t\minus
t^\prime]$). Upon making the ansatz that our order parameters are
smooth functions of time, we then lose the remaining microscopic
variables $\{z(t)\}$ in the retarded self-interaction term (which
are automatically converted into averages over their distribution)
and end up with the effective single trader problem
\begin{eqnarray}
  \frac{d}{dt}q(t) &=& \tilde{\theta}(t)
  -\alpha\int_{0}^ t\!dt^\prime~R(t,t^\prime)
  \bra \sigma[q(t^\prime),z]\ket_z+\sqrt{\alpha}\eta(t)
  \label{eq:singleagent_conttime}
\end{eqnarray}
The correlation- and response functions
(\ref{eq:meaningof_C},\ref{eq:meaningof_K}) are the dynamic order
parameters of the problem, and must now be solved
self-consistently from the following closed equations
\begin{eqnarray}
  C(t,t^\prime)
&=&
  \left\bra ~\bra
    \sigma[q(t),z]\ket_z~\bra\sigma[q(t^\prime),z]\ket_z~
  \right\ket_\star
  \label{eq:finalC}
\\[1mm]
  G(t, t^\prime)
&=&
  \frac{\delta }{\delta \tilde{\theta}({t^\prime})}
  \left\bra~
    \bra\sigma[q(t),z]\ket_z~
  \right\ket_\star \label{eq:finalG}
\end{eqnarray}
(for $t\neq t^\prime$). The brackets $\bra \ldots\ket_\star$ now
refer to the stochastic process (\ref{eq:singleagent_conttime})
(which no longer involves the $\{z(t)\}$, only their
distribution). Note that the correlation- and response functions
are generally discontinuous at $t=t^\prime$, viz. $C(t,t)=1$ and
$G(t,t)=0$ (in our derivation we have used the It\^{o}
convention). For additive noise with
$P(z)=\frac{1}{2}K[1-\tanh^2(Kz)]$ (as in e.g.
\cite{ChalMarsZecc00,CavaGarrGiarSher99}, with $K$ such that
$\int\!dz~P(z)z^2=1$), for instance, one has
$\bra\sigma[q,z]\ket_z=\tanh[\beta q]$ with inverse `temperature'
$\beta\equiv K/T$. While for multiplicative noise one finds:
$\bra\sigma[q,z]\ket_z=\lambda(T) \sgn(q)$ with $\lambda(T)=\int\!
dz\, P(z)\sgn[1+Tz]$.

What remains is to take the continuous time limit in  our
expressions for the effective noise covariance and the retarded
self-interaction kernel. The latter, defined by
(\ref{eq:finishedR}), simply becomes
\begin{eqnarray}
R(t,t^\prime)&=&\lim_{\delta\to 0}\frac{1}{\delta}[\openone+\delta
G]^{-1}_{t t^\prime} \nonumber
\\
 &=&
\delta(t-t^\prime)+\sum_{\ell> 0}(-1)^\ell G^\ell(t,t^\prime)
\label{eq:final_exact_R}
\end{eqnarray}
with the usual definition for multiplication of the continuous
time kernels  $G^{\ell+1}(t,t^\prime)=\int\!dt^\pprime
~G^\ell(t,t^\pprime)G(t^\pprime,t^\prime)$. The continuous time
limit of the  noise covariance kernel (\ref{eq:final_sigma})
becomes
\begin{eqnarray}
  \Sigma(s_0,s_0^\prime)
&=&
  \sum_{\ell \ell^\prime\geq 0}(-1)^{\ell+\ell^\prime}
  \lim_{\delta\to 0}
  \delta^{\ell+\ell^{\prime}}\!\!\!\!\!\!
  \sum_{s_0>\ldots >s_\ell \geq 0}
  \sum_{s_0^\prime >\ldots >s_\ell^\prime \geq 0}
  \prod_{i=0}^{\ell} \prod_{j=0}^{\ell^\prime}\left[1+\delta_{s_i
  s^\prime_j}\frac{\rate}{2\delta}\right]
  \nonumber
  \\
&&
  \times
  [1\!+\!C(s_\ell,s_\ell^\prime)] G(s_0, s_1) \ldots  G(s_{\ell-1}, s_\ell)
  G(s_0^\prime, s^\prime_1) \ldots  G(s^\prime_{\ell^\prime-1},
  s_\ell^\prime)
  \nonumber\\
&=&
  \sum_{\ell \ell^\prime\geq 0}
  (-1)^{\ell+\ell^\prime} \!\int_0^\infty \!ds_1 \ldots ds_\ell
  ds_1^\prime \ldots ds_\ell^\prime \prod_{i=0}^{\ell}
  \prod_{j=0}^{\ell^\prime}
  \left[1+\frac{1}{2}\rate\delta(s_i-s^\prime_j)\right]
  \nonumber
  \\
&&
  \times
  [1\!+\!C(s_\ell,s_\ell^\prime)] G(s_0, s_1) \ldots  G(s_{\ell-1}, s_\ell)
  G(s_0^\prime, s^\prime_1) \ldots  G(s^\prime_{\ell^\prime-1},
  s_\ell^\prime)
  \nonumber\\
&&
  \label{eq:final_exact_sigma}
\end{eqnarray}

\section{Stationary State in the Ergodic Regime}

In this section we study the long time limit of the effective
single agent process (\ref{eq:singleagent_conttime}) in the
 regime where the following three conditions are met: time translation
invariance, i.e.
\be
    \lim_{t\rightarrow\infty} C(t+\tau,t) =
 C(\tau),\qquad
  \lim_{t\rightarrow\infty} G(t+\tau,t) = G(\tau),
  \label{eq:ass1}
\ee a  finite integrated response (or static susceptibility), i.e.
\be
 \lim_{t\rightarrow\infty} \int\!dt' G(t,t') = \chi < \infty
 \label{eq:ass2}
\ee and weak long-term memory \cite{CuglKurc95},
i.e.
\be
 \lim_{t\rightarrow\infty} \int_0^{t_w}\!dt'\,G(t,t')=0\quad
\mbox{for any fixed}~~t_w
\label{eq:ass3}
 \ee
Together these three conditions ensure that also the retarded
self-interaction $R$ and the noise covariance matrix $\Sigma$ will
become time translation invariant:
$\lim_{t\rightarrow\infty}R(t+\tau,t)=R(\tau)$ and
$\lim_{t\rightarrow\infty}\Sigma(t+\tau,t)=\Sigma(\tau)$.  We
introduce the following notation for long time averages:
$\overline{f}=\lim_{\tau\rightarrow\infty} \tau^{-1}\int_0^{\tau
}dt\,f(t)$. Upon assuming the above three conditions to hold, we
can calculate the long time average for the single agent process
(\ref{eq:singleagent_conttime}), giving:
\begin{equation} \label{eq:singleagentss}
  \overline{dq/dt}
  =
  \overline{\theta}
  - \frac{\alpha}{1+\chi}\overline{\bra \sigma \ket_z}
  + \sqrt{\alpha}~\overline{\eta}.
\end{equation}
It has been noted and exploited  earlier (in e.g.
\cite{ChalMars99} and \cite{HeimCool01}) that agents can be
divided into two categories: frozen agents who get completely
fixed
 on one specific strategy, and fickle agents who always continue to
alternate their strategies. The first type of agent will have a
non-zero average preference velocity, $\overline{dq/dt}\ne 0$,
while for the agents in the latter group one has
$\overline{dq/dt}=0$. For fickle agents it follows from equation
(\ref{eq:singleagentss}) that $\overline{\bra \sigma\ket_z}=
(1+\chi)(\overline{\theta}+\sqrt{\alpha}\overline{\eta})/\alpha$.
This implies that a necessary condition for an agent to be fickle
is:
\begin{equation}\label{eq:ficklecons}
{\rm fickle~agents:}~~~~~~~~~~
 |\overline{\theta}+\sqrt{\alpha}\overline{\eta}|
   \leq
     \frac{\alpha}{1+\chi} |\bra \sigma[\pm\infty,z]\ket_z | \equiv \gamma.
\end{equation}
Note that $|\bra \sigma[\pm \infty,z]\ket_z|=\lambda(T)$ for
multiplicative noise, and that $|\bra \sigma[\pm
\infty,z]\ket_z|=1$ for additive noise. With the conventions
$\lambda=\lambda(T)$ for multiplicative noise and $\lambda=1$ for
additive noise, we cover both cases by writing $\gamma=\lambda
\alpha/(1+\chi)$. If an agent is frozen, the preference velocity
$\overline{dq/dt}$ must have the same sign as $q(t)$, and thus
also the same sign as $\overline{ \bra \sigma \ket_z}$. This
implies
\begin{equation}\label{eq:frozencons}
{\rm frozen~agents:}~~~~~~~~~~
  |\overline{\theta}+\sqrt{\alpha}\overline{\eta}|\geq
    \frac{\alpha}{1+\chi} |\overline{\bra \sigma \ket_z}|
  =\gamma
\end{equation}
Since the two conditions (\ref{eq:ficklecons},\ref{eq:frozencons})
are complementary, they are are not only necessary but also
sufficient for characterizing agents as either fickle
(\ref{eq:ficklecons}) or frozen (\ref{eq:frozencons}). The
asymptotic behaviour of the agent is thus completely determined by
the persistent noise $\overline{\eta}$, which is a Gaussian random
variable with zero mean and a variance given by
\begin{equation}
  \bra \overline{\eta}^2 \ket_\star
  =
  \lim_{t\rightarrow \infty}
  \frac{1}{t}\int_0^t dt'  \, \Sigma(t')
  =
  \frac{1+c}{(1+\chi)^2},
\end{equation}
in which we find the persistent auto-correlation $c=
\lim_{\tau\rightarrow \infty}\tau^{-1}\int_0^\tau dt  \, C(t)$.
The latter observable can be expressed in terms of the integrated
response $\chi$, the market's control parameter $\alpha$, and the
fraction of frozen agents $\phi$. To do this we first separate the
expression for $c$ into frozen and fickle contributions, by using
the conditions (\ref{eq:ficklecons},\ref{eq:frozencons}) (we may
set $\overline{\theta}=0$):
\begin{eqnarray}
  c
&=&
  \left\bra
    \overline{\bra \sigma\ket_z}^2
    \Theta(|\sqrt{\alpha}\overline{\eta}|-\gamma)
  \right\ket_\star
  +
  \left\bra
    \overline{\bra \sigma\ket_z}^2
    \Theta(\gamma-|\sqrt{\alpha}\overline{\eta}|)
  \right\ket_\star
\nonumber \\
&=&
  \lambda^2 \left\bra
     \Theta(|\sqrt{\alpha}\overline{\eta}|-\gamma)
  \right\ket_\star
  +\left(\frac{1+\chi}{\sqrt{\alpha}}\right)^2
   \left \bra
     \overline{\eta}^2
     \Theta(\gamma-|\sqrt{\alpha}\overline{\eta}|)
  \right\ket_\star,
\end{eqnarray}
where $\Theta$ denotes the step-function. Only Gaussian integrals
remain (defining the distribution of $\overline{\eta}$). To
compactify the final result it is convenient to introduce
\begin{equation}
\label{eq:y}
  y=\frac{\lambda\sqrt{\alpha}}{\sqrt{2(1+c)}}.
\end{equation}
The fraction of frozen agents $\phi$ and the persistent
correlation $c$ can now be written as
\begin{eqnarray}
\phi&=& 1-\erf(y) \label{eq:phiselfconsistent}
\\
  c
&=&
  \lambda^2\left(
   \phi(y)
    +
   \frac{2}{y^2} \int_{0}^y \frac{dx}{\sqrt{\pi}}\,x^2 e^{-x^2}
  \right).
  \label{eq:cselfconsistent}
\end{eqnarray}
Under the conditions (\ref{eq:ass1},\ref{eq:ass2},\ref{eq:ass3})
we can calculate the static susceptibility without knowing the
non-persistent parts of the response-  and correlation functions.
The reason for this is that, asymptotically, the frozen agents
will not change their preferences when subjected to an
infinitesimal external perturbation field, whereas the fickle
agents will react linearly to such a field. In formulae:
\[
  \chi
=
  \lim_{t\rightarrow\infty} \int_0^t \!dt'\,
  \frac{\delta}{\delta \tilde{\theta}(t')}
  \left\bra \bra \sigma[q(t),z]\ket_z\right\ket_\star
=
  \frac{\partial}{\partial \overline{\theta}}
  \left\bra \overline{\bra \sigma \ket_z} \right\ket_\star
=
  \frac{1}{\alpha}(1-\phi)(1+\chi).
\]
Or, equivalently:
\begin{equation} \label{eq:chiselfconsistent}
  \chi=\frac{1-\phi}{\alpha-(1-\phi)}.
\end{equation}
The equations (\ref{eq:y}), (\ref{eq:cselfconsistent}) and
(\ref{eq:chiselfconsistent}) completely specify the stationary
behaviour of the system. They are exact for the on-line minority
game. We will not give a detailed exposition of the behaviour of
the solutions of the equations here, because they have been
presented and discussed at length in earlier papers. The
interested reader is referred to \cite{MarsChalZecc00} and
\cite{HeimCool01}  in particular.

However, it is appropriate at this point to discuss the status  of
our assumptions (\ref{eq:ass1},\ref{eq:ass2},\ref{eq:ass3}).
Equation (\ref{eq:chiselfconsistent}) tells us that $\chi$ is
positive and finite for $\alpha>1-\phi$, and will diverge at
$\alpha= 1-\phi$. Numerical solution of
(\ref{eq:y},\ref{eq:cselfconsistent},\ref{eq:chiselfconsistent})
shows that this happens at $\alpha=\alpha_c(T)$, where
$\alpha_c(0)\approx 0.3374$, and $\alpha_c(T)$ tends to $0$ as
$T\to \infty$ (see \cite{CoolHeimSher01} for the phase diagram in
the $(\alpha,T)$ plane). For $\alpha>\alpha_c(T)$ simulations and
theory are found to be in perfect agreement, and there is no
evidence that the assumptions
(\ref{eq:ass1},\ref{eq:ass2},\ref{eq:ass3}) do not hold. Below
$\alpha_c(T)$, however, the system's behaviour is known to depend
strongly on initial conditions
\cite{GarrMoroSher00,HeimCool01,MarsChal01b}. This indicates that
most likely not only condition (\ref{eq:ass2}) (finite integrated
response) ceases to hold, but also (\ref{eq:ass3}) (weak long-term
memory). These two conditions may seem very similar; however, work
in progress on a version of the MG where agents try to correct for
their own impact on the market \cite{DemaMars00} indicates that it
is possible to have a long-term memory \emph{and} a finite
integrated response \cite{HeimDema01}. For the original minority
game as presented here we have found no evidence that this can
happen. Hence, equations (\ref{eq:cselfconsistent}) and
(\ref{eq:chiselfconsistent}) can for $\alpha>\alpha_c(T)$ be
regarded as an exact and complete description of the persistent
order parameters of the minority game.

\section{The Volatility}

\subsection{Definitions and Exact Relations}

An important measure of the efficiency of a market is the average
mismatch of buyers and sellers. In the case of the minority game
this is given by the magnitude of the total bid $A(\ell)$. Since
in the present model the long term average
$\overline{A}=\lim_{L\rightarrow\infty}\frac{1}{L}\sum_{\ell\leq
L} A(\ell)$ vanishes, the appropriate measure here is the
volatility $\sigma$, which measures the size of the fluctuations
of $A$ (see equation (\ref{eq:Amu})) in the stationary state:
\begin{equation}
  \sigma^2
  =
  \lim_{L\rightarrow\infty}\frac{1}{L}\sum_{\ell=1}^L A(\ell)^2
  =
  \lim_{L\rightarrow\infty}\frac{1}{L}\sum_{\ell=1}^L
    \{A^{\mu(\ell)}[\bq(\ell),\bz(\ell)]\}^2.
\label{eq:define_true_vola}
\end{equation}
If the limit $L\rightarrow\infty$ is taken before the
thermodynamic limit $N\to\infty$, then the volatility will be
self-averaging with respect to the realization of the presentation
of the patterns $\{\mu(\ell)\}$, with respect to the realization
of the decision noise $\{\bz(\ell)\}$, and with respect to the
realization of the quenched disorder variables $\bOmega$ and
$\bxi$. Hence, in the thermodynamic limit, the average volatility
 (over the above sources of randomness) will be identical to the single
sample volatility. In the continuous-time version of the process,
i.e. after the introduction of Poisson-distributed iteration
durations, one may write:
\begin{equation}
  \sigma^2=
  \lim_{\tau\rightarrow\infty}
  \frac{1}{\tau}\int_0^\tau\!dt\, \left[
    \frac{1}{p}\sum_\mu
      \left\bra
         \bra A^\mu[\bq(t),\bz(t)]^2\ket_\bz
      \right\ket
  \right]_{dis},
\label{eq:define_vola}
\end{equation}
Here the average without subscripts refers to the full stochastic
process (including both the randomness in the selection of the
external  information as well as that induced by the decision
noise). The approach followed in the present paper in fact allows
us to study a more general object than the volatility:
\begin{equation}
  \Xi(t,t^\prime)
  =
  \frac{1}{p}\sum_\mu \left[
     \left\bra
        \bra A^\mu[\bq(t),\bz(t)] A^\mu[\bq(t'),\bz(t')]\ket_\bz
     \right\ket
  \right]_{dis},
  \label{eq:define_volamatrix}
\end{equation}
From (\ref{eq:define_volamatrix}) the volatility follows as
\be
\sigma^2=
  \lim_{\tau\rightarrow\infty}
  \frac{1}{\tau}\int_0^\tau\!dt~\Xi(t,t)
  \label{eq:sigma_in_Xi}
\ee
 In contrast to (\ref{eq:define_vola}), the quantity
(\ref{eq:define_volamatrix}) also describes transient bid
fluctuations and their correlations. The quantity
(\ref{eq:define_volamatrix}) is calculated via a simple adaptation
of the corresponding calculation in \cite{HeimCool01} (to which we
refer for full details). One introduces the observables
$A^\mu_t=\Omega^\mu+x_t^\mu/\sqrt{2}$ explicitly into the
calculation of the generating functional using a delta function
(written in integral form, which generates the conjugate variables
$\hA^\mu_t$), similar to the introduction of  $x^\mu_t$ and
$w^\mu_t$.  After having done the integral over $\Omega^\mu$ one
then  finds $A^\mu_t=(u+x_t)/\sqrt{2}$. If we apply this
modification to equations
(\ref{eq:Phi_workingout1},\ref{eq:Poissonterm}), and perform the
Gaussian integrals over $u$, $x$, $\hx$, $w$, $\hw$ and $\hA$ (in
this order) we are left with (modulo an irrelevant constant):
\begin{eqnarray}
  \Phi
&=&
  \frac{1}{N}\sum_\mu \log
\left\bra \int \D A^\mu
   \exp\left[
     -\sum_{tt'}A^\mu_t \left[
       (\openone+GE)^\dag D^{-1} (\openone+GE)
     \right]_{tt'}A^\mu_{t'}
   \right]\right\ket_n
   \nonumber
\end{eqnarray} From this it follows that the covariance matrix
(\ref{eq:define_vola}) is half the inverse of the matrix appearing
in the above exponential between $A^\mu_t$ and $A^\mu_{t^\prime}$.
\begin{equation} \label{eq:Xi}
  \Xi_{tt^\prime}
  =
  \frac{1}{2}
  \left\bra
     \left[(\openone+GE)^{-1}D(\openone+EG^\dag)^{-1}\right]_{tt^\prime}
  \right\ket_n
\end{equation}
In this expression one can carry out explicitly the average over
the variables $\{n_t\}$, similar to the calculation of
 the noise covariance matrix (\ref{eq:final_exact_sigma}) (to which
 (\ref{eq:define_volamatrix}) is found to be similar but, in contrast to the
  batch cases studied in
  \cite{HeimCool01,CoolHeimSher01},
 not proportional).
Upon subsequently taking the continuous time limit $\delta\to 0$
this leads to an exact expression for
 the generalized volatility matrix (\ref{eq:define_volamatrix}),
 at any combination of times $(t,t^\prime)$, in terms of our
 dynamical order parameters $C(t,t^\prime)$ and $G(t,t^\prime)$.

\subsection{Expression for the Volatility in Terms of Persistent Order Parameters}

Since in practice it is very difficult to solve the order
parameter equations (\ref{eq:finalC},\ref{eq:finalG}) for finite
temporal separations, it would be helpful to find an  expression
for (\ref{eq:Xi}) (and hence also the volatility) which involves
persistent order parameters only. In the remainder of this section
we discuss procedures to achieve this in the ergodic regime
$\alpha>\alpha_c(T)$. First we turn to the long time correlation
in bids. If $t$ and $t^\prime$ are sufficiently separated and
conditions (\ref{eq:ass1},\ref{eq:ass2},\ref{eq:ass3}) hold (so
that $G$ decays effectively on finite time-scales), the Poisson
average in (\ref{eq:Xi}) factorizes over the two $E$ matrices, so
that in the continuous-time limit the long term bid correlations
reduce to
\begin{eqnarray}
  \Xi(\infty)&=&\lim_{\tau\rightarrow\infty}
  \lim_{t\rightarrow\infty}
  \Xi(t+\tau,t)
  \nonumber\\
  &=&\lim_{\delta\rightarrow 0}
  \lim_{\tau\rightarrow\infty}
  \lim_{t\rightarrow\infty}
  \frac{1}{2}
  \left[
    (\openone+\delta G)^{-1}D(\openone+ \delta G^\dag)^{-1}
  \right]_{t+\tau,t}
  \nonumber\\
  &=& \frac{1}{2} \frac{1+c}{(1+\chi)^2}
\end{eqnarray}
This result is exact. For the volatility itself, which in view of
(\ref{eq:ass1},\ref{eq:ass2},\ref{eq:ass3}) can be written as
$\sigma^2=\lim_{t\rightarrow\infty}\Xi(t,t)$, and which depends on
the detailed short-time structure of the kernels $G$ and $D$, one
has to resort to approximations. One such approximation is
motivated by a property of mean-field disordered systems with
detailed balance, where fluctuation-dissipation theorems (FDT)
allow one to gauge away the non-persistent parts of the response
and correlation function while leaving averages containing a
single time index and averages containing infinitely separated
times unchanged (see e.g. \cite{FisherHertz}); for equilibrium
systems the resulting simpler equations are exact. Although no
general analogons of equilibrium FDT's are as yet known for
non-equilibrium systems such as the minority game, one could
assume that the resulting recipe for removing non-persistent
contributions to the various kernels also applies to
non-equilibrium systems; this as yet ad-hoc assumption has
recently been applied with remarkable success to a similar
non-equilibrium problem  \cite{HeimCool01b}. Here it would amount
to  removing all non-persistent parts of the various kernels,
while retaining the relations $\delta\sum_t G_{t t^\prime}=\chi$,
$\frac{1}{\tau}\sum_{t\leq \tau}C_{t,t^\prime}=c$ and $C_{tt}=1$,
i.e. one inserts
\begin{equation}
  C_{t,t^\prime}=c+(1-c)\delta_{tt^\prime}, \qquad
  G_{t,t^\prime}=\chi \gamma (1-\gamma \delta)^{t-t^\prime-1}~~~(t>t^\prime),
\end{equation}
and takes $\gamma\rightarrow 0$ at the end of the calculation. The
only expressions containing $G$ which will survive this limit, are
those were each instance of $G$ is summed over the whole history.
More specifically, in the volatility
\begin{equation}
  \sigma^2=
  \frac{1}{2}\sum_{\ell\geq 0}\sum_{\ell^\prime\geq 0}
  \left\bra
    \left[
      (GE)^{\ell} D (EG^\dag)^{\ell^\prime}
    \right](0)
  \right\ket_n
  \label{eq:FDTlike}
\end{equation}
contractions like $\delta \sum_s G_{t s}G_{t^\prime s}$ cannot
survive, as they are of order $\gamma$. Hence the Poisson average
in (\ref{eq:FDTlike}) factorizes over the $E$ matrices, and the
only non-vanishing term involving the diagonal part of $D$ is the
term $\ell=\ell^\prime=0$ term. The result\footnote{Note: in the
batch case \cite{HeimCool01,CoolHeimSher01} one finds a slightly
different expression.} is
\begin{equation}
  \sigma^2
  \approx
  \frac{1}{2}\frac{1+c}{(1+\chi)^2}
  + \frac{1}{2}(1-c).
  \label{eq:approx_vola}
\end{equation}
This expression for the volatility depends only on persistent
order parameters, and is independent of the learning rate $\rate$.
Work is in progress to explore and understand the theoretical
basis, if it exists, of (or disprove the correctness of, as the
case may be) the elimination of the non-persistent parts in
general mean-field disordered systems without detailed balance,
which underlies the derivation of (\ref{eq:approx_vola}) from the
exact expression (\ref{eq:FDTlike}) in the ergodic regime.

\section{Relation Between Present Exact Solution and Previous Work}

The present study improves upon and generalizes previous work on
the on-line MG in several ways: it is exact for $N\to\infty$, it
is a dynamical theory (with statics included as a spin-off), and
it deals with a large family of decision noise definitions
(including additive and multiplicative noise, and intermediates).
In this section we compare the various approximations and
assumptions made in previous studies with the exact solution, and
also compare the solution of the on-line MG with that of the batch
MG.

\subsection{Expressions for the Diffusion Matrix}

The first study in which one finds an expression for the diffusion
matrix of the microscopic on-line process is \cite{GarrMoroSher00}
(concerned with both additive and multiplicative noise).
Comparison with the exact expressions
 (\ref{eq:diffusion_A},\ref{eq:diffusion_B}) shows that
the matrix given in \cite{GarrMoroSher00} can be regarded as an
approximation obtained by disregarding fluctuations due to the
selection of external information, and retaining only those
generated by the decision noise. Note that, similar to
(\ref{eq:diffusion_B}), it contains the factors $(1-\tanh^2[\beta
q_i])$ and vanishes as $\beta\to\infty$. The diffusion term
presented more recently (for additive noise)
 in \cite{MarsChal01b} can also be seen as an
approximation of  (\ref{eq:diffusion_A},\ref{eq:diffusion_B}),
obtained upon replacing $\sum_{\mu} \xi_k^\mu\xi_\ell^\mu \bra
A^\mu[\bq,\bz]^2 \ket_\bz $ by $\lim_{\tau\rightarrow
\infty}\tau^{-1}\sum_{t\leq \tau}\sum_{\mu} \xi_k^\mu\xi_\ell^\mu
\frac{1}{p}\sum_\nu \bra A^\nu[\bq(t),\bz]^2 \ket_\bz $, i.e.  by
neglecting certain correlations and subsequently introducing an
additional temporal average over $\bq$.

Note that neither of the two studies
\cite{GarrMoroSher00,MarsChal01b} attempt to solve the dynamics;
the objective of \cite{GarrMoroSher00} was to highlight the
presence and role of the diffusion term in the microscopic
equations (since at the time the microscopic laws were claimed to
be effectively deterministic \cite{ChalMarsZecc00}). The authors
of \cite{MarsChal01b} have to  restrict themselves to statics,
since in their approximation and subsequent analysis they neglect
the time dependence of the fluctuation term which they replace by
the volatility (or, equivalently, they do not regard their
volatility as time-dependent). However, their approach does shed
additional light on the nature of the phase transition at
$\alpha=\alpha_c(T)$ and the system state for
$\alpha<\alpha_c(T)$.

\subsection{The Impact of Truncating the Kramers-Moyal Expansion}

Initially, in \cite{ChalMarsZecc00}, the KM expansion was
truncated after the Liouville term, leading to deterministic
microscopic laws. In \cite{GarrMoroSher00,MarsChal01b} it was
truncated after the Fokker-Planck term. The question we turn to
now is: what is the effect of such truncations, assuming one would
have taken the correct diffusion matrix (i.e. both terms
(\ref{eq:diffusion_A}) and (\ref{eq:diffusion_B})) ? This question
is even more relevant in view of the fact that, for additive
noise, the exact equations
(\ref{eq:y},\ref{eq:cselfconsistent},\ref{eq:chiselfconsistent}),
which give  the stationary solution in the ergodic regime,  are
identical to those found earlier in \cite{ChalMarsZecc00}.

If we trace back our derivations, we find that the effect of
truncating the KM expansion after the deterministic term or after
the Fokker-Planck term  is to replace the function $\phi$ in
 (\ref{eq:Poissonterm}) by
\begin{eqnarray}
  \phi_{\rm Det}[\{x,w\},u]
&=&
  e^{\delta \sum_t i w_t(u+x_t)},
\end{eqnarray}
or by
\begin{eqnarray}
\phi_{\rm FP}[\{x,w\},u]
  &=&
  e^{
     \delta \sum_t i   w_t(u+x_t)
     -\frac{1}{4}\rate \delta \sum_t
        w^2_t(u+x_t)^2 },
\end{eqnarray}
respectively. It turns out that both expressions can be written in
the  form (\ref{eq:Poissonterm}), i.e.
\begin{equation}
  \phi[\{x,w\},u]
=
  \left\bra e^{\frac{1}{2}i\rate \sum_t n_t w_t[u+x_t]}
  \right\ket_n,
\end{equation}
\begin{equation}
  \bra f[n_1,n_2,\ldots] \ket_{n}
=
  \sum_{n_1,n_2\ldots \geq 0}
  \left[
    \prod_s P_{2\delta/\rate}[n_s]~f[n_1,n_2,\ldots]
  \right]
\end{equation}
but with alternative definitions for the statistics of the random
variables $\{n_t\}$:
\begin{eqnarray}
 {\rm Deterministic}: &~~~& P_a[n]=\delta[n-a]
 \\
 {\rm Fokker\!-\!Planck}: &~~~& P_a[n]=\frac{e^{-\frac{1}{2}(n-a)^2/a}}{\sqrt{2\pi
 a}}
\end{eqnarray}
Both truncations can thus be seen as approximations of the true
(Poisson) distribution which describes the noise due to random
information selection: in the deterministic case one replaces the
true  $P_{2\delta/\rate}[n]$ by a delta-distribution with the
correct first moment $\bra n\ket=2\delta/\rate$,
 in the
Fokker-Planck case one replaces the true $P_{2\delta/\rate}[n]$ by
a Gaussian distribution and ensures that the first {\em two}
moments $\bra n\ket=2\delta/\rate$ and $\bra
n^2\ket=(2\delta/\rate)^2 +2\delta/\rate$ are correct. The above
representation allows us to continue with our original derivation,
in spite of the truncations, right up to and including the
equations describing the effective single trader process
(\ref{eq:singleagent_conttime},\ref{eq:finalC},\ref{eq:finalG}),
since the choice made for $P_a[n]$ only affects the kernels
$R(t,t^\prime)$ and $ \Sigma(t,t^\prime)$. The derivation of
expression (\ref{eq:final_exact_R}), however, only involved the
first moment of $P_a[n]$ (see section 6), so {\em both}
truncations would have led to the exact expression for
$R(t,t^\prime)$. Surprisingly, the derivation of
(\ref{eq:final_exact_sigma}) only involved the first two moments
of $P_a[n]$, so whereas the deterministic truncation would have
led to an incorrect expression for the covariance matrix
$\Sigma(t,t^\prime)$ (obtained by putting $\rate\to 0$ in
(\ref{eq:final_exact_sigma})), the Fokker-Planck approximation
 would have led to the correct
expression (\ref{eq:final_exact_sigma}) and hence to the exact
dynamical order parameter equations.

This explains why such truncations, although not a priori
justified, here can lead to correct results. The deterministic
theory would lead at most to correct equations in ergodic
stationary states (where the learning rate of
(\ref{eq:final_exact_sigma}) drops out), but would fail to
describe dynamical properties. The Fokker-Planck truncation would
lead to the correct macroscopic dynamical theory, provided one
uses the correct diffusion matrix, because the relevant order
parameters of the present model fortunately turn out not to be
sensitive to the (weak) non-Gaussian fluctuations.

\subsection{Approximations of the Volatility}

Since the volatility (\ref{eq:define_true_vola}) involves
non-persistent order parameters (describing correlations over
short temporal separations, even in the stationary state), it
could not be calculated directly in any of the previous studies.
Instead one had to resort to approximations.

We repeat an argument here which was
 first given in \cite{ChalMars99}, and which applies in the ergodic $\alpha>\alpha_c(T)$
 regime. Upon assuming that phase space is sampled ergodically by the process,
one may replace the time and pattern presentation averages by an
average over the equilibrium probability measure:
\begin{equation}
  \sigma^2=
  \left[
    \left\bra
       \frac{1}{p}\sum_\mu
         \bra A^\mu[\bq,\bz]^2\ket_\bz
      \right\ket
  \right]_{dis},
\end{equation}
The decision noise average factorizes over sites, i.e. $\bra
\sigma[q_i,z_i] \sigma[q_j,z_j] \ket_\bz= \bra \sigma[q_i,z_i]
\ket_\bz \bra \sigma[q_j,z_j] \ket_\bz$ for $i\ne j$  (this is not
an approximation). It was subsequently argued that in ergodic
states the mean field character of the system means that also the
ensemble average factorizes over sites, i.e. $\bra \sigma[q_i,z_i]
\sigma[q_j,z_j] \ket= \bra \sigma[q_i,z_i] \ket \bra
\sigma[q_j,z_j] \ket $ for $i\ne j$. This  should be regarded as
an approximation, since, although for large $N$ non-diagonal
correlations can be assumed weak, they are also $N$ in number and
therefore cannot simply be discarded.
 The result of this mean field approximation is:
\begin{eqnarray}
  \sigma^2
&\approx&
  \frac{1}{p}\sum_\mu \left[
      \left\bra \bra A^\mu \ket_\bz  \right\ket^2
  \right]_{dis}
  +
  \frac{1}{p}\sum_\mu \frac{1}{N}\sum_i \left[
     \xi_i^\mu \xi_i^\mu \left\{
      1- \left\bra \bra \sigma[q_i,z] \ket_z  \right\ket^2
    \right\}
  \right]_{dis},\nonumber\\
&=&\label{eq:meanfieldvolatility}
  \frac{1}{p}\sum_\mu \left[
      \left\bra \bra A^\mu \ket_\bz  \right\ket^2
  \right]_{dis}
  +
  \frac{1}{2}  \left[
      1- \frac{1}{N}\sum_i \left\bra \bra \sigma[q_i,z] \ket_z  \right\ket^2
  \right]_{dis} + \order({N^{-1/2}})
\end{eqnarray}
which then leads, in the ergodic regime and for additive decision
noise, more or less directly to
\begin{equation} \label{eq:mfsigma}
 \sigma^2
\approx
  \frac{1}{2}\frac{1+c}{(1+\chi)^2}
  + \frac{1}{2}(1-c),
\label{eq:approx_vola_again}
\end{equation}
which is equation (\ref{eq:approx_vola}). This identification
shows that the exactness or otherwise of the (traditional)
approximation (\ref{eq:approx_vola_again}) is crucially linked to
the question of under which conditions non-persistent parts of
dynamic order parameters can be `transformed away' in ergodic
non-equilibrium models (see section 9). It also emphasizes that
such approximations must fail in the $\alpha<\alpha_c(T)$ regime,
since in equilibrium systems such procedures are based on
fluctuation-dissipation theorems and therefore typically reproduce
the replica-symmetric solution.

Numerical evidence presented in
\cite{ChalMarsZecc00,MarsChalZecc00} shows that
(\ref{eq:approx_vola_again}) is a very good approximation in the
$\alpha>\alpha_c(T)$ region, until just above the transition point
$\alpha_c(T)$. It is not clear whether the slight discrepancy just
above $\alpha_c(T)$ is due to insufficient equilibration in the
simulation, or because the approximation breaks down.

\subsection{Relation with the Dynamical Solution of the Batch MG}

Finally, comparison shows that the present dynamical solution
(\ref{eq:singleagent_conttime},\ref{eq:finalC},\ref{eq:finalG})
for the on-line MG can be regarded as a straightforward
continuous-time equivalent of the discrete-time dynamical solution
of the batch MG \cite{HeimCool01,CoolHeimSher01}, obtained simply
by substituting time derivatives for discrete differences and
integral kernels for matrices, for any type of decision noise. The
only real difference is in the occurrence of the learning rate
$\rate$ in (\ref{eq:final_exact_sigma}) (which makes sense, since
it reflects fluctuations relating to the random choice of external
information, which are absent by definition in the batch models),
which will be responsible for differences between batch and
on-line MG models in the transients and in the non-ergodic region.
This learning rate term, however, does not affect the stationary
state solution for $\alpha>\alpha_c(T)$, which explains why the
stationary state of the batch models
\cite{HeimCool01,CoolHeimSher01} and the locations $\alpha_c(T)$
of their ergodicity-breaking transitions
 were identical to those found earlier (by others) for
 the on-line MG.

\section{Discussion}

In this paper we have solved the dynamics of the on-line
 minority game (MG), with general types of decision noise (including additive and multiplicative
decision noise as specific choices). We have done so by following
the successful approach which also recently
 led to the exact solution of discrete-time batch versions of the MG
\cite{HeimCool01,CoolHeimSher01} (i.e. the application of
 generating functional techniques a la De Dominicis \cite{DeDo78}), and
 we have dealt with the  problems relating to temporal regularization,
 which occur only in on-line MG's, using the (exact) procedure of
 \cite{BedeLakaShul71}.
The end result is a macroscopic dynamical  theory  in the form of
closed equations for correlation- and response functions (the
dynamical order parameters of the problem) which are defined via
an effective continuous-time single-trader process. These
equations are exact for $N\to\infty$ (where $N$ denotes the number
traders), and they incorporate all static and dynamic properties
of the on-line MG,
 both in the
ergodic and in the non-ergodic regime.

We have used our theory to resolve a number of open problems
related to approximations and assumptions made in previous
studies. For instance, we show why it is in principle not allowed
to truncate the Kramers-Moyal expansion of the microscopic process
after the Fokker-Planck term (let alone after the flow term), but
why upon doing so one can for the present version of the MG  still
find the correct macroscopic equations, we confirm that the
different diffusion matrices for the Fokker-Planck term in the
process as proposed earlier by others are incomplete or
approximate, and we indicate how previously proposed
approximations involving the market volatility can be traced back
to assumptions relating to ergodicity.

The macroscopic theory now available is not only exact, but also
more comprehensive than its on-line predecessors: it deals with
the full dynamics (with statics included as a by-product), and it
generalizes the class of decision noise definitions. We also hope
that our theory will end the discussions about which are the
correct microscopic equations for the MG, and that it can be used
in future as the canonical starting point for both analyses and
generalizations (e.g. dynamics in the non-ergodic regime, or using
the real market history as external information as originally
proposed in \cite{ChalZhan97}),
 and for approximations (for which there is still a
need, since it is generally a non-trivial task to solve our
macroscopic laws). This might well include some of the
approximations which have already been proposed earlier, which can
now be provided with transparent interpretations and with a guide
for systematic improvement.

\section*{Acknowledgements}

JAFH would like to thank the King's College London Association for
support. ACCC and JAFH would like to thank the organizers of the
`International Seminar on Statistical Mechanics of Information
Processing in Cooperative Systems' (Dresden, 2001) where many
fruitful discussions helped shaping the present study.



\section*{References}

\begin{thebibliography}{99}

\bibitem{ChalZhan97}
Challet D and  Zhang Y-C 1997 \emph{Physica A} {\bf 246}  407

\bibitem{Arth94}
Arthur W B 1994 \emph{Am. Econ. Assoc. Papers and Proc.} {\bf 84}  406

\bibitem{Chalweb}
Challet D, {\tt http://www.unifr.ch/econophysics/minority}

\bibitem{Cava99}
Cavagna A 1999 {\em Phys. Rev. E} {\bf 59}  R3783

\bibitem{SaviManuRiol99}
Savit R, Manuca R and Riolo R 1999 \emph{Phys. Rev. Lett.} {\bf 82}  2203

\bibitem{CavaGarrGiarSher99}
Cavagna A, Garrahan J P, Giardina I, and Sherrington D 1999 {\em
Phys. Rev. Lett.} {\bf 83}  4429

\bibitem{Johnson1}
Hart M, Jefferies P, Johnson N F and Hui P M 2001 {\em Physica A}
(in press)

\bibitem{Johnson2}
Hart M, Jefferies P and Johnson N F 2000 {\em Phys. Rev. E} {\bf
63} 017102

\bibitem{Johnson3}
Jefferies P, Hart M, Johnson N F and Hui P M 2000 {\em J. Phys A:
Math. Gen.} {\bf 33} L409


\bibitem{ChalMars99}
Challet D and Marsili M 1999 {\em Phys. Rev. E} {\bf 60} 6271

\bibitem{ChalMarsZecc00}
Challet D, Marsili M and Zecchina R 2000 {\em Phys. Rev. Lett.}
{\bf 84} 1824

\bibitem{MarsChalZecc00}
Marsili M, Challet D and Zecchina R 2000 {\em Physica A} {\bf 280}  522

\bibitem{CommTrieste}
Challet D, Marsili M and Zecchina R 2000 {\em Phys. Rev. Lett.}
{\bf 85} 5008

\bibitem{CommOxford}
Cavagna A, Garrahan J P, Giardina I and Sherrington D 2000 {\em
Phys. Rev. Lett.} {\bf 85} 5009

\bibitem{GarrMoroSher00}
Garrahan J P, Moro E and Sherrington D 2000 {\em Phys. Rev E} {\bf
62} R9

\bibitem{MarsChal01a}
Marsili M and Challet D 2001 {\em Adv. in Complex Sys.}{\bf 3}-I

\bibitem{MarsChal01b}
Marsili M and Challet D 2001  preprint {\tt cond-mat/0102257}

\bibitem{BedeLakaShul71}
Bedeaux D, Lakatos-Lindenberg K and Shuler K 1971 {\em J. Math.
Phys.} {\bf 12} 2116

\bibitem{DeDo78}
De Dominicis C 1978 {\em Phys. Rev. B} {\bf 18}  4913

\bibitem{HeimCool01}
Heimel J A F and Coolen A C C 2001 {\em Phys. Rev. E} {\bf 63}
056121

\bibitem{CoolHeimSher01}
Coolen A C C, Heimel J A F and Sherrington D 2001 preprint {\tt
cond-mat/0106635}

\bibitem{CuglKurc95}
Cugliandolo L F and Kurchan J 1995 {\em Phil. Mag. B} {\bf 71} 501

\bibitem{DemaMars00} De Martino A and Marsili M 2000 {\em J. Phys. A}
{\bf 34} 2525-2537

\bibitem{HeimDema01} Heimel J A F and De Martino A  2001 {\em in preparation}

\bibitem{HeimCool01b}
Heimel J A F and Coolen A C C 2001 preprint {\tt cond-mat/0102272}

\bibitem{FisherHertz}
Fisher K H and Hertz J A 1991 {\em Spin Glasses} (Cambridge: U.P.)


\end{thebibliography}
\end{document}